\documentclass[preprint,11pt]{elsarticle}

\usepackage[a4paper]{geometry}
\usepackage{amssymb}
\usepackage{amsmath}
\usepackage{amsmath,bm}
\usepackage{xcolor}
\usepackage{caption}
\usepackage{subcaption}
\usepackage{multirow}
\usepackage{graphicx,import}
\usepackage{siunitx}
\graphicspath{{Pictures_pdf/}}
\usepackage{lineno}

\setcitestyle{numbers}
\journal{Journal of Manufacturing Processes}

\begin{document}

\begin{frontmatter}



\title{Enhancing multiscale simulations for spark plasma sintering with a novel Direct FE$^2$ framework}

\author[label1]{A. Kumar\corref{cor1}}
\author[label1]{Z. Zhang}
\author[label1]{M. Bambach}
\author[label1]{M. Afrasiabi}

\affiliation[label1]{organization={Advanced Manufacturing Lab, ETH Zurich},
            addressline={Leonhardstrasse 21}, 
            city={Zurich},
            postcode={8092}, 
            country={Switzerland}}


\cortext[cor1]{Corresponding author}
\fntext[label2]{E-mail address: askumar@ethz.ch }


\begin{abstract}
The spark plasma sintering (SPS) process, a key technology for advanced material manufacturing, demands accurate and efficient simulation tools to capture the complex electro-thermal-mechanical interactions inherent in powder materials. This paper introduces a novel concurrent multiscale framework employing the Direct FE$^2$ method, designed for fully coupled electro-thermal-mechanical simulations in SPS. The model integrates microscale powder characteristics into a macroscopic analysis through multi-point constraints within a 3D finite element (FE) solver. This approach enables, for the first time, a direct and seamless coupling of micro- and macroscale physical phenomena, enhancing both accuracy and computational efficiency by capturing interactions across scales. The proposed method achieves a temperature and displacement error margin below 1\% compared to full FE analysis while reducing computational degrees of freedom by a factor of 8, resulting in a 70-fold acceleration in simulation time. Additionally, the methodology provides robust flexibility in accommodating diverse powder morphologies without compromising precision, enabling degree-of-freedom reductions of up to 44 times. This combination of enhanced efficiency and accuracy establishes the proposed Direct FE$^2$ approach as a highly effective tool for realistic and scalable simulations of the SPS process.
\end{abstract}




\end{frontmatter}


\section{Introduction}
\label{sec:introduction}
Sintering is a manufacturing process in which powdered materials are compacted within a die and heated to elevated temperatures, just below their melting point, to form a dense component. This method offers several advantages, including high material efficiency, the ability to produce complex shapes, and precise control over the porosity of the final product. It is widely used in industries such as automotive, aerospace, and electronics. Among the various sintering techniques, spark plasma sintering (SPS) is often the method of choice due to its rapid densification and reduced processing time, which are achieved through the application of electric current and pressure \cite{grasso2009electric, hu2020review}. Optimizing critical parameters such as temperature, pressure, holding time, and powder characteristics (e.g., size, distribution, shape, and composition) is crucial to controlling the final product's density and microstructure. Therefore, developing a numerical model that accurately incorporates the powder material’s characteristics is essential, as it has a significant influence on the properties of the final product.

In modeling SPS, the powder can be treated either as a continuum or as discrete particles. Continuum approaches, such as those using finite element methods (FEM), are numerically efficient and sufficiently accurate for macroscale phenomena like thermal analysis. However, they fall short in capturing detailed microscale effects at the powder level \cite{olevsky2012fundamental, kraft2004numerical, kumar2023two}. For example, these methods may overlook crucial insights into localized particle deformation during sintering and its influence on the densification process and material properties. To address this, continuum mechanics approaches require a sintering model that accurately represents densification behavior, such as Olevsky’s model \cite{olevsky2012fundamental} or Abouf's model \cite{abouaf1988finite}, which incorporate various material transport mechanisms to predict densification. Nonetheless, these models lack particle-level information, such as particle size distribution, shape, and particle-particle interactions. Although grain boundary and thermal diffusion effects have been included in extended models by Olevsky and Froyen \cite{olevsky2009impact}, these models are rarely used in process modeling due to the complexity of identifying constitutive material and diffusion parameters, as well as the absence of direct particle-particle interaction data.

Alternatively, when powder particles are treated as discrete entities, the Discrete Element Method (DEM) is often preferred due to its ability to represent the thermomechanical interactions between individual particles \cite{martin2014simulation}. DEM allows for a detailed representation of particle-particle contact, accounting for key factors such as contact forces, friction, heat transfer, and particle rearrangement during sintering. This enables a clearer understanding of microscale phenomena, such as localized stress distributions and particle deformation, which are critical to the densification process and the resulting material properties. However, despite its advantages, DEM faces challenges in simulating large plastic deformations due to limitations in its contact models, which restrict its application primarily to the initial stages of particle compaction (at low densities).

Recently, Ransing et al. \cite{ransing2000powder} and Cameron and Gethin \cite{cameron2001exploration, gethin2001numerical} have explored a hybrid approach that integrates finite element methods (FEM) and discrete element methods (DEM), known as the Multi-Particle Finite Element Method (MPFEM). This approach effectively combines the discrete characteristics of individual particles with the continuous bulk properties of the powder mass, making MPFEM well-suited for modeling solid-state sintering \cite{wang2022numerical} and hot pressing \cite{xu2021mpfem, zou2020investigation}. MPFEM has been widely applied in the cold compaction of metal powders, including copper \cite{zhang2015multi, guner2015numerical}, aluminum \cite{kyung2009densification}, and iron \cite{gustafsson2013multi}, as well as in composite powders such as Fe-Al \cite{han2018mpfem} and Al-SiC \cite{huang2017multi}. Despite its efficiency, MPFEM is predominantly used for relatively simple physical problems and has yet to be applied to the complex multiphysics (electro-thermal-mechanical) interactions encountered in SPS. Additionally, the computational limitations of MPFEM restrict the number of particles in the simulation, typically to around 100. This constraint limits the scope of simulations to small-scale geometries, often in the micrometer range. Furthermore, MPFEM uses explicit time integration schemes that require extremely small time steps, resulting in very large computational times.

In contrast, the FE$^2$ method is a multiscale computational technique that simultaneously performs FE simulations at two separate scales: one at the macroscale level of the overall structure and another at the microscale level of the representative volume element (RVE) \cite{feyel1999multiscale, feyel2000fe2}. The macroscale analysis evaluates structural deformations, while the microscale analysis of RVEs determines the stress-strain state at each integration point within the macroscale elements. Consequently, the material properties at the microscale RVE level are incorporated into the macroscale computations. This method has been extensively developed for analyzing the characteristics of heterogeneous materials.

This method combines micro- and macroscale models within a single simulation framework, coupling them through multi-point constraints in a consistent finite element analysis, thus offering high computational efficiency \cite{lange2021efficient, raju2021review}. By incorporating microscale RVEs to capture powder characteristics, the Direct FE$^2$ method significantly reduces the total number of particle interactions compared to MPFEM. This reduction in computational complexity improves the robustness of the method for addressing multiphysics interaction problems. Furthermore, the Direct FE$^2$ approach can be easily integrated with both commercial and open-source FE solvers via multi-point constraints, which has facilitated its broader application in various fields, including static and dynamic analysis \cite{zhi2021transient, zhi2022direct}, structural analysis \cite{xu2022direct, yeoh2022multiscale}, composite damage \cite{raju2021analysis}, and thermal analysis \cite{zhi2023multiscale, meng2024direct}. Although this method is commonly used in the simulation of composite materials, its application to powder processing simulations remains unexplored and is introduced in this paper.

The literature review on SPS process simulation reveals the following research gaps:
\begin{itemize}
    \item Although the SPS process is inherently an electro-thermal-mechanical problem, only a limited number of studies have addressed this multiphysics coupling. However, these studies overlook the powder characteristics at the particle level.
    
    \item The majority of SPS modeling frameworks treat the powder as a continuum, typically using FEM, which does not capture the detailed information at the powder level.
    
    \item Simulation methods such as MPFEM and DEM, which integrate powder property information, are seldom applied in SPS simulations due to the high computational demands and the complexity introduced by numerous particle interactions, as well as the multiphysics nature of the problem.
\end{itemize}

To overcome these limitations, we propose a novel 3D multiscale modeling framework employing the Direct FE$^2$ method that concurrently integrates the full interactions at the powder level (microscale) with the electro-thermal-mechanical effects at the part scale (macroscale). This extension is used to simulate powder compaction and sintering behavior. Furthermore, we demonstrate the accuracy and computational efficiency of applying the Direct FE$^2$ method to SPS process simulations.

This manuscript is structured as follows. Section 2 presents the theoretical background and modeling framework of the coupled Direct FE$^2$ method. Section 3 presents a numerical verification case that compares the proposed method with the full FE model. Section 4 explores the application of this method to powder compaction and sintering, including its validation, along with simulations of various powder morphologies and SPS geometry. In conclusion, Section 5 presents the findings of this research and outlines future directions for further investigation.


\begin{figure}
\renewcommand{\arraystretch}{0.90}
	\centering
	\small		
	\begin{tabular}{|l l  l l |}
		\hline
        \multicolumn{2}{|l}{\textbf{Nomenclature}} & & \\
        $\vec{b}$ & Body force vector &  $\boldsymbol\sigma $ & Cauchy stress tensor \\
       $\vec{J}$ & Current density &  $\rho$ & Density \\        
        $ u $ & Displacement &    $\boldsymbol{\varepsilon}^{e}$ & Elastic strain \\
  $\vec{E}$ & Electric field intensity      &  $V$ & Electric potential \\        
         $\dot{\varepsilon}_{eq}$ & Equivalent strain rate & $ \sigma_{eq}$ & Equivalent stress   \\
         $\xi$ & Gauss quadrature point & $C_{p}$ & Heat capacity \\
       $\vec{q}$ & Heat flux vector  & 	$r$ & Heat generated \\
        $\boldsymbol{\varepsilon}^{p}$ & Plastic strain & $\Bar{\alpha}_{\xi}$ & RVE scale factor  \\
        $\dot{\boldsymbol{\varepsilon}}$ & Strain rate tensor & $\vec{t}$ & Traction force vector \\
         $T$ & Temperature  & $\boldsymbol{k_{c}}$ & Thermal conductivity tensor \\
          $\boldsymbol{\varepsilon}^{th}$ & Thermal strain &  & \\	  
		\hline
	\end{tabular}
\end{figure}

\section{Multiphysics and multiscale modeling framework}
SPS is a complex process that integrates multiple interacting physical phenomena, necessitating a comprehensive multiphysics modeling framework. The process uses electric current to generate heat through the Joule heating effect, while external pressure is applied to accelerate material densification. This interplay establishes a tightly coupled electro-thermal-mechanical system. The electrical and thermal phenomena are intricately linked. Temperature-dependent electrical conductivity affects the distribution of current, which in turn determines internal heat generation. This heat propagates through the material via conduction and is stored within it, contributing to the overall thermal response.

Simultaneously, the mechanical behavior is intertwined with the thermal and electrical domains. Temperature-dependent material properties, such as elastic modulus and thermal expansion, directly influence the material's deformation under load. The inelastic deformation of the material also generates heat, creating a feedback loop between mechanical deformation and thermal dynamics. These inter-dependencies underscore the need for a multiphysics approach that accounts for the interactions between electrical current, heat generation and transfer, and mechanical deformation.

In addition to the multiphysics considerations, a multiscale approach is essential for accurately modeling SPS processes. This is necessary because the phenomena governing the behavior of materials during sintering occur at various scales, from powder-powder interactions (i.e., microscale) to macroscopic densification behavior. In this section, the governing equations are presented with sufficient detail for computer implementation, followed by a description of the concurrent multiscale modeling procedure and its implementation details in our selected commercial FE code, Abaqus.

\subsection{Governing equations}
The weak form of the static mechanical balance equation, derived from the principle of virtual work, can be expressed as
\begin{equation}
    \int_{\Omega} \delta \boldsymbol{\varepsilon} \mathrel{:}  \boldsymbol{\sigma} d\Omega = \int_{\Omega} \delta u \vec{b} d\Omega + \int_{\Gamma} \delta u \vec{t} d\Gamma
\end{equation}
where $\boldsymbol{\varepsilon}$ represents the strain tensor, $\boldsymbol{\sigma}$ is the Cauchy stress tensor, $u$ denotes displacement, $\vec{t}$ is the traction force vector, $\vec{b}$ is the body force vector, and $\Omega$ and $\Gamma$ denote the domain and its boundary, respectively. The internal virtual work associated with the mechanical balance equation is given by:
\begin{equation}
\label{eq:virtual_work_M}
    \delta W_{int}^M =  \int_{\Omega} \delta\boldsymbol{\varepsilon} \mathrel{:} \boldsymbol{\sigma} d \Omega
\end{equation}
The weak form of the thermal energy balance equation can be written as:
\begin{equation}
\label{eq:thermal_weak_form}
    \int_{\Omega} \rho C_p \dot{T} \delta T d \Omega + \int_{\Omega} \nabla(\delta T) \cdot \boldsymbol{ k}_c \cdot \nabla T d \Omega = \int_{\Omega} \delta T r d \Omega + \int_{\Gamma} \delta T q^* d \Gamma
\end{equation}
where $\rho$ is the density, $C_p$ the heat capacity, $T$ the temperature, $\boldsymbol{k_c}$ the thermal conductivity matrix, and $q^*$ the heat flux per unit area. For the electric-thermal problem, heat is generated solely by Joule heating (with no external heat sources), represented as $r = \vec{J} \cdot \vec{E}$, where $\vec{J}$ is the current density and $\vec{E}$ is the intensity of the electric field, expressed in terms of the electric potential $V$ as
\begin{equation}
    \vec{E} = - \nabla V
\end{equation}
Similarly, by applying the virtual work to Maxwell's charge conservation equation, the weak form can be expressed as
\begin{equation}
\label{eq:electrical_weak_form}
      \int_{\Omega} \nabla \delta V \cdot \boldsymbol{\sigma}_c \cdot \nabla V d \Omega= \int_{\Omega} \delta V r_c d \Omega + \int_{\Gamma} \delta V J d \Gamma
\end{equation}
Here, $r_c$ represents the internal volumetric current source per unit volume, and $\boldsymbol{\sigma}_c$ is the electrical conductivity matrix. Eq. \eqref{eq:thermal_weak_form} and Eq. \eqref{eq:electrical_weak_form} are used in the coupled electrical and thermal analysis. The internal virtual work is expressed as
\begin{equation}
\label{eq:virtual_work_ET_0}
    \delta W_{int}^{ET} =  \int_{\Omega} \nabla \delta V \cdot \boldsymbol{\sigma}_c\cdot \nabla V d\Omega + \int_{\Omega} \rho C_p \dot{T} \delta T d\Omega  + \int_{\Omega} \nabla \delta T \cdot k_c \cdot \nabla T d\Omega
\end{equation}
By employing the expressions $ \vec{q}  =  - \boldsymbol{k_c} \cdot \nabla T$ and $\vec{J} = \boldsymbol{\sigma}_c \cdot \vec{E}$, Eq. \eqref{eq:virtual_work_ET_0} can be rewritten as
\begin{equation}
\label{eq:virtual_work_ET}
    \delta W_{int}^{ET} =  \int_{\Omega} \nabla \delta V \cdot \vec{J} d\Omega + \int_{\Omega} \rho C_p \dot{T} \delta T d\Omega  + \int_{\Omega} \nabla \delta T \cdot \vec{q} d\Omega
\end{equation}
By combining equation Eq. \eqref{eq:virtual_work_M} and Eq. \eqref{eq:virtual_work_ET}, the total internal virtual work of the coupled electro-thermal-mechanical system is expressed as:
\begin{equation}
\label{eq:total_virtual_work}
    \delta W_{int} =   \int_{\Omega} \delta \boldsymbol{\varepsilon} \mathrel{:} \boldsymbol{\sigma} d \Omega + \int_{\Omega} \nabla \delta \Omega \cdot \vec{J} d \Omega + \int_{\Omega} \rho C_p \dot{T} \delta T d \Omega  + \int_{\Omega} \nabla \delta T \cdot \vec{q} d\Omega
\end{equation}
After applying the Gaussian quadrature for numerical integration in finite element analysis, Eq. \eqref{eq:total_virtual_work} becomes:
\begin{equation}
\label{eq:total_energy_gauss}
     \delta W_{int} = \sum^{n_e}_{e=1} \sum^{n_{\xi}}_{\xi=1} \alpha_\xi J_\xi ( \delta \boldsymbol{\varepsilon}  \mathrel{:} \boldsymbol{\sigma} )_{\xi}  +  \sum^{n_e}_{e=1} \sum^{n_{\xi}}_{\xi=1} \alpha_\xi J_\xi ( \nabla \delta V  \cdot \vec{J} + \rho C_p \dot{T} \delta T   +  \nabla \delta T \cdot \vec{q}  \big )_{\xi}
\end{equation}
where $\xi$ denotes a Gauss quadrature point in element $e$, $n_{\xi}$ is integration point in each element, and $n_e$ is total number of elements. $J_{\xi}$ and $\alpha_{\xi}$ are the Jacobian and the weight for the Gauss point, respectively. 

\subsection{Concurrent multiscale modeling and implementation procedure}
The Direct FE$^2$ approach begins by identifying an RVE that captures the material's microstructure. In conventional FE$^2$, the objective is to solve the equations at the macroscale FE level using the average values of stress, heat flux, and current flux concurrently determined at the microscale (RVE) level. The Direct FE$^2$ approach integrates these two levels of nested FE computations into a single FE analysis through geometric scaling and kinematic constraints.

To distinguish between the micro- and macroscale, variables with a tilde ($\Tilde{\Box}$) are evaluated at the microscale, while unmarked variables are evaluated at the macroscale. The same convention is applied to the gradients of these variables. Therefore, Eq.\eqref{eq:total_energy_gauss} becomes:
\begin{equation}
    \begin{split}   
      \delta W_{int} = & \sum^{n_e}_{e=1} \sum^{n_{\xi}}_{\xi=1} \alpha_\xi J_\xi \langle \delta \Tilde{\boldsymbol{\varepsilon}} \rangle_{\xi}  \mathrel{:} \langle \Tilde{\boldsymbol{\sigma}} \rangle_{\xi}  +  \sum^{n_e}_{e=1} \sum^{n_{\xi}}_{\xi=1} \alpha_\xi J_\xi \langle \nabla \delta \Tilde {V} \rangle_{\xi}  \cdot \langle \vec{\Tilde J} \rangle_{\xi} + \\
   &   \langle \Tilde \rho \Tilde {C}_p \rangle_{\xi} \langle \dot{\Tilde{T}} \rangle_{\xi} \langle \delta \Tilde  T \rangle_{\xi}   +  \langle \nabla \delta \Tilde T \rangle_{\xi} \cdot \langle\vec{\Tilde{q}} \rangle_{\xi}
    \end{split}
\end{equation}

\begin{equation}
\label{eq:virtual_work_GP}
      \delta W_{int} = \sum^{n_e}_{e=1} \sum^{n_{gp}}_{\xi=1} \alpha_\xi J_\xi ( \delta \boldsymbol{\varepsilon}  \mathrel{:} \boldsymbol{\sigma} )_{\xi}  +  \sum^{n_e}_{e=1} \sum^{n_{gp}}_{\xi=1} \alpha_\xi J_\xi ( \nabla \delta V  \cdot \vec{J} + \rho C_p \dot{T} \delta T   +  \nabla \delta T \cdot \vec{q}  \big )_{\xi}
\end{equation}
The operator $\langle * \rangle_{\xi}$ denotes volume-averaged quantities over the RVE associated with Gauss point $\xi$, defined as:
\begin{equation}
    \langle * \rangle = \frac{1}{|\Tilde{\Omega}|} \int_{\Tilde{\Omega}} * d \Tilde{\Omega}
\end{equation}
where $|\Tilde{\Omega}|$ represents the volume of the microscale RVE. The micro-to-macro transition is governed by the Hill-Mandel condition, which ensures energy consistency between the macroscale and microscale:
\begin{equation}
\label{eq:HM_mech}
   \underbrace{\langle \delta \Tilde{\boldsymbol{\varepsilon}} \rangle  \mathrel{:} \langle \Tilde{\boldsymbol{\sigma}} \rangle}_{\text{macro}} = \underbrace{\langle \delta \Tilde{\boldsymbol{\varepsilon}} \mathrel{:} \Tilde{\boldsymbol{\sigma}} \rangle}_{\text{micro}} = \frac{1}{|\Tilde{\Omega}|} \int_{\Tilde{\Omega}} \delta \Tilde{\boldsymbol{\varepsilon}} \mathrel{:} \Tilde{\boldsymbol{\sigma}}  d \Tilde{\Omega}
\end{equation}
\begin{equation}
\label{eq:HM_current}
     \langle \nabla \delta \Tilde {V} \rangle  \cdot \langle \vec{\Tilde J} \rangle = \langle \nabla \delta \Tilde {V}   \cdot \vec{\Tilde J} \rangle =  \frac{1}{|\Tilde{\Omega}|} \int_{\Tilde{\Omega}} \nabla \delta \Tilde {V}   \cdot \vec{\Tilde J} d \Tilde{\Omega}
\end{equation}
\begin{equation}
\label{eq:HM_temp}
    \langle \Tilde \rho \Tilde {C}_p \rangle \langle \dot{\Tilde{T}} \rangle \langle \delta \Tilde  T \rangle   +  \langle \nabla \delta \Tilde T \rangle \cdot \langle\vec{\Tilde{q}} \rangle =\langle \Tilde \rho \Tilde {C}_p \dot{\Tilde{T}}  \delta \Tilde  T    + \nabla \delta \Tilde T  \cdot\vec{\Tilde{q}} \rangle = \frac{1}{|\Tilde{\Omega}|} \int_{\Tilde{\Omega}}  \Tilde \rho \Tilde {C}_p \dot{\Tilde{T}}  \delta \Tilde  T    + \nabla \delta \Tilde T  \cdot\vec{\Tilde{q}}  d \Tilde{\Omega}
\end{equation}
By Combining Eq. \eqref{eq:HM_mech}-\eqref{eq:HM_temp} into Eq. \eqref{eq:virtual_work_GP}, the total sum of internal virtual work can be expressed as:
\begin{equation}
\label{eq:total_internal_work}
 \begin{split}  
    \delta W_{int} & = \sum^{n_e}_{e=1} \sum^{n_{\xi}}_{\xi=1} \alpha_\xi J_\xi  \langle \delta \Tilde{W}_{int} \rangle _{\xi} \\
    & = \sum^{n_e}_{e=1} \sum^{n_{\xi}}_{\xi=1} \alpha_\xi J_\xi \frac{1}{|\Tilde{\Omega}|} \int_{\Tilde\Omega} \big ( \delta \Tilde{\boldsymbol{\varepsilon}} \mathrel{:} \Tilde{\boldsymbol{\sigma}}  + \nabla \delta \Tilde{V} \cdot \vec{\Tilde{J}}  + \Tilde{\rho} \Tilde{C}_p \dot{\Tilde{T}} \delta \Tilde{T} +  \nabla \delta \Tilde{T} \cdot \vec{\Tilde{q}} \big ) d\Tilde{\Omega}
    \end{split}
\end{equation}
The total internal virtual work from microscale FE analyses of all RVEs is given by:
\begin{equation}
\label{eq:RVE_total_internal_work}
     \delta \Tilde{W}_{int} =  \sum^{n_e}_{e=1} \sum^{n_{\xi}}_{\xi=1}\int_{\Tilde\Omega} \big ( \delta \Tilde{\boldsymbol{\varepsilon}} \mathrel{:} \Tilde{\boldsymbol{\sigma}}  + \nabla \delta \Tilde{V} \cdot \vec{\Tilde{J}}  + \Tilde{\rho} \Tilde{C}_p \dot{\Tilde{T}} \delta \Tilde{T} +  \nabla \delta \Tilde{T} \cdot \vec{\Tilde{q}} \big ) d\Tilde{\Omega}
\end{equation}
Comparing Eqs. \eqref{eq:total_internal_work} and \eqref{eq:RVE_total_internal_work}, it is evident that $\delta W_{int}$ is equivalent to a scaled sum of $\delta \Tilde{W}_{int}$, with each RVE scaled by the factor:
\begin{equation}
\label{eq:Scalling_factor}
    \Bar{\alpha}_{\xi} =  \frac{\alpha_\xi J_\xi}{|\Tilde{\Omega}|}
\end{equation}
To ensure energy equilibrium, the volume of the microscale RVE at each Gauss point must be scaled. In 2D problems, this is achieved by adjusting the thickness, while in 3D, RVEs are scaled equally in all three dimensions \cite{tan2020direct}. The internal energy at the macroscale is replaced by contributions from the microscale, resulting in a single analysis that incorporates both macro and micro degrees of freedom.

In this study, the RVEs are scaled by proportionally adjusting their dimensions in all three directions. The scaling factor for each RVE is determined based on the volume of the corresponding integration point, as described in Eq. \eqref{eq:Scalling_factor}.

The RVEs at different integration point are not directly connected, but they are weakly coupled via the macro nodes. This coupling between scales is akin to the macro-to-micro transition in conventional FE$^2$ methods, which apply macroscale kinematics -- such as temperature, displacement, and potential gradients -- to define the RVE boundary conditions. There are three common boundary conditions that satisfy the Hill-Mandel condition:
\begin{itemize}
\item Linear displacement boundary conditions $\rightarrow$ Voig-Taylor model

\item Traction or constant stress boundary conditions $\rightarrow$ Reuss-Hill model

\item Periodic boundary conditions
\end{itemize}
We use periodic boundary conditions for the numerical simulations in this work, as they are typically preferred due to their good approximation performance and symmetry properties \cite{tan2020direct,meng2024direct}. These conditions are expressed as
\begin{equation}
\label{eq:pbc_T}
    \Tilde{T}^+ - \Tilde{T}^- = (N_I (x^+) - N_I (x^-) ) T_I 
\end{equation}
\begin{equation}
    \Tilde{\boldsymbol{u}}^+ - \Tilde{\boldsymbol{u}}^- = (N_I (x^+) - N_I (x^-) ) \boldsymbol{u}_I
\end{equation}
\begin{equation}
    \Tilde{V}^+ - \Tilde{V}^- = (N_I (x^+) - N_I (x^-))V_I
\end{equation}
Here, $N_I$, $T_I$, $V_I$, and $\boldsymbol{u}_I$ represent the shape function, temperature, voltage, and displacement of the macroscale finite element node $I$, where the RVE resides. The coordinate system for the 3D RVE is centered at the origin ($x_0$), with the plus (+) and minus (-) signs indicating points on opposing RVE edges or faces.

Additional conditions must be imposed to constrain rigid body translation of the RVEs. One node in each RVE, typically the center, is tied to the corresponding quadrature point of the macroscale element. To avoid overall displacement between the RVE and macroscale element, temperature and potential constraints are applied at the central node as follows:
 \begin{equation}
     \Tilde{\boldsymbol{u}}_0 = N_I(x_0) \boldsymbol{u}_I
 \end{equation}
 \begin{equation}
     \Tilde{T}_0 = N_I(x_0) T_I
 \end{equation}
 \begin{equation}
 \label{eq:pbc_V0}
     \Tilde{V}_0 = N_I(x_0) V_I
 \end{equation}

In ABAQUS, Eqs. \eqref{eq:pbc_T} to \eqref{eq:pbc_V0} are directly enforced through multi-point constraints (MPCs), which establish a connection between the macroscale and microscale degrees of freedom \cite{abel1979algorithm}. These MPCs are implemented in ABAQUS using the constraint equation method (keyword *Equation).

\begin{figure}[h]
    \centering
    \includegraphics[width=0.85\linewidth]{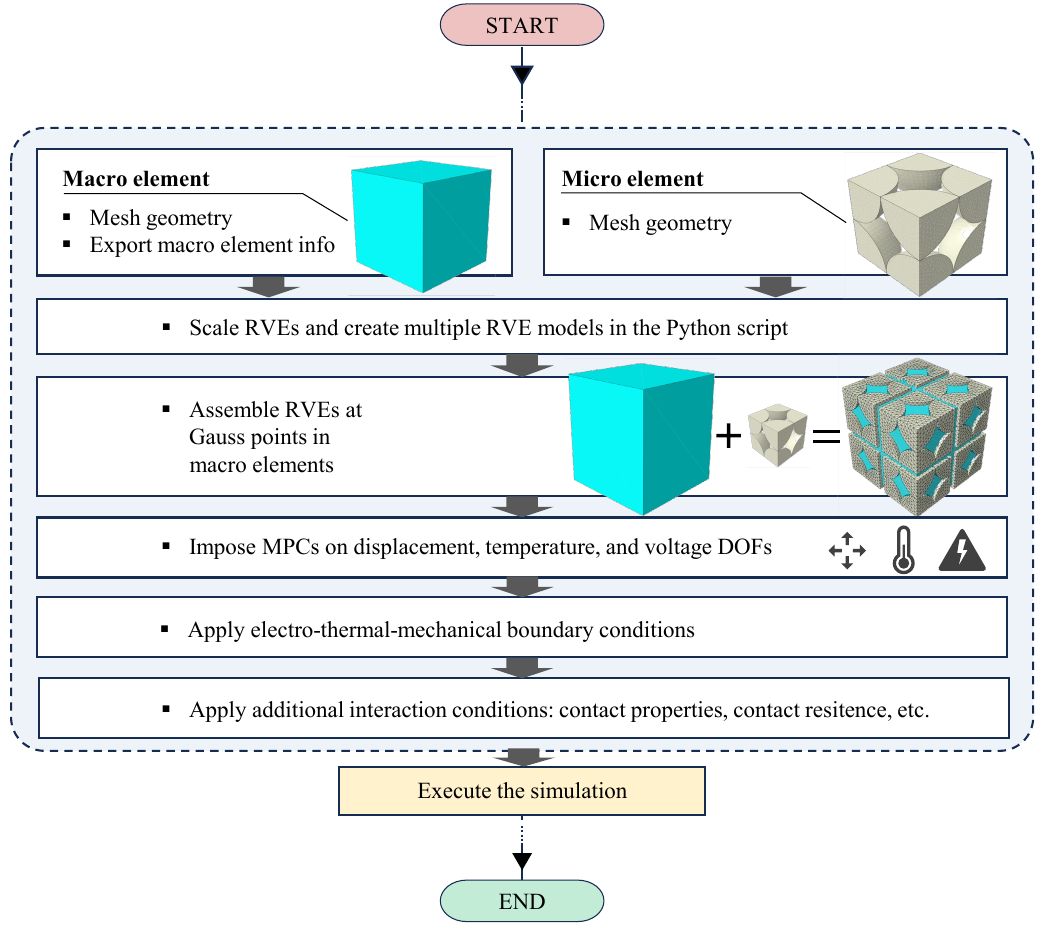}
    \caption{Flow chart of the implementation process of the Direct FE$^2$ method.}
    \label{fig:Flow_Chart}
\end{figure}

The Direct FE$^2$ method for fully coupled electro-thermo-mechanical simulations is implemented within the commercial FE solver Abaqus. This implementation leverages Abaqus’s support for linear MPCs and coupled electro-thermo-mechanical analysis via the \textit{Coupled Electrical-Temperature-Displacement} feature. The process, illustrated in Fig. \ref{fig:Flow_Chart}, involves the following steps:
\begin{itemize}
     \item Develop macro and micro RVE models and mesh the geometry in the macro domain ($\Omega$) and the micro RVE domain ($\Tilde{\Omega}$) using the appropriate element type (e.g., Abaqus element Q3D8 for coupled electro-thermal-mechanical elements).
     
     \item Replicate multiple RVE instances and align their centers with the Gauss points (integration points) of the macro elements. Each RVE instance is scaled according to Eq. \eqref{eq:Scalling_factor} in all dimensions.
     
     \item Assign dummy material properties (such as near-zero stiffness, density, thermal conductivity, and specific heat) to macro elements, which serve to enforce boundary conditions, while actual material properties are assigned to the RVEs.
     
     \item Implement MPC equations to enforce periodic boundary conditions between macro- and microelements, integrating both models into a unified FE model via Abaqus by imposing kinematic constraints through MPCs.
\end{itemize}

\section{Verification}
The implementation of electro-thermal-mechanical coupled physics in the Direct FE$^2$ method is verified by comparing its results with a full FE simulation. The comparison is performed on a cuboid geometry, and the results are evaluated in terms of displacement and temperature. The validation of the proposed method is presented in Section \ref{sec:validation}. For mechanical and thermal scenarios, the Direct FE$^2$ method has been validated by Tan et al. \cite{tan2020direct} and Zhi et al. \cite{zhi2023multiscale}  respectively.

\begin{figure}
    \centering
    \includegraphics[width=\linewidth]{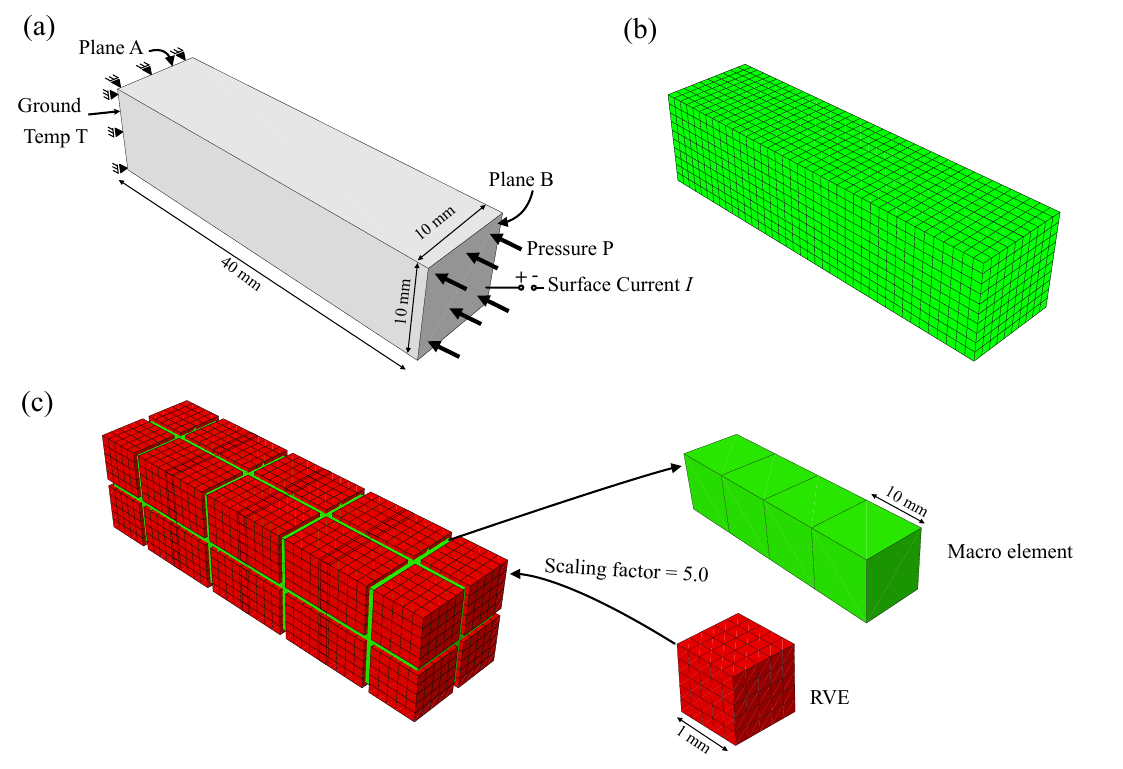}
    \caption{Schematic of (a) geometry and boundary conditions considered for the verification test and meshed geometry used for (b) full FE and (c) Direct FE$^2$ method.}
    \label{fig:SG_Geometry}
\end{figure}

The cuboid has dimensions of 10$\times$10$\times$40 mm, as shown in Fig. \ref{fig:SG_Geometry}a. Figs \ref{fig:SG_Geometry}b and \ref{fig:SG_Geometry}c show the meshed geometry used for the full FE and the Direct FE$^2$ method, respectively. For the full FE model, a mesh size of 1 mm is used, and an electro-thermal-mechanical coupled element (Q3D8) is applied. The Direct FE$^2$ model, shown in Fig. \ref{fig:SG_Geometry}c, consists of a macro model with RVEs at each integration point. The macro element mesh size is set to 10 mm, resulting in four macro elements for this case. Each RVE is 1$\times$1$\times$1 mm with a mesh size of 0.2 mm. The scaling factor for the RVE is set to 5.0, as given by Eq. \eqref{eq:Scalling_factor}.

The electro-thermal properties of the material used in this study are listed in Table \ref{table:1_copper_mat_properties_ET}. An elastoplastic material model, with parameters provided in Table \ref{table:1_copper_mat_properties_Elasto_Plastic}, is used. The material has a Young's modulus of 115 GPa and a Poisson's ratio of 0.34 \cite{loidolt2018modeling}. While specific material properties are assigned to the full FE model and the RVEs in the Direct FE$^2$ model, the macro elements in the Direct FE$^2$ model are assigned dummy material properties.

To simplify the verification study, boundary conditions are applied as shown in Fig. \ref{fig:SG_Geometry}a. A pressure of 200 MPa and a surface current flux of 100 A/mm$^2$ are applied to one end of the cuboid (plane B), while the opposite end (plane A) is grounded and constrained in the perpendicular direction. One point is fixed to prevent rigid body motion, and a temperature of 300 K is applied to plane A.

\begin{table}
\centering
 \renewcommand*{\arraystretch}{1.0}
\small
\caption{Electro-thermal materials properties of copper \cite{anselmi2005fundamental}}
\begin{tabular}{l l l} 
 \hline 
  \textbf{Parameter} & \textbf{Unit} & \textbf{Expression}  \\ 
 \hline 
   $C_p $ & \unit{J/(kg\,K)} & 355.3 + 0.1$T$\\ 
   $k_c $ & \unit{W/(m\,K)}  & 420.66 + 0.07$T$ \\ 
  $\rho_{c} $ & \unit{\Omega m } & $(5.5 + 0.038T) \times 10^{-9}$\\ 
  $\rho $ & \unit{ kg/m^{3}} & 8960\\ 
\hline 
\end{tabular}
\label{table:1_copper_mat_properties_ET}
\end{table}

\begin{table}
\centering
 \renewcommand*{\arraystretch}{1.0}
\small
\caption{Material properties of copper used as particle material \cite{loidolt2018modeling}}
\begin{tabular}{l l l l l l l l} 
 \hline 
  \textbf{Parameter} & \textbf{Unit} & \multicolumn{6}{c}{\textbf{Value}}  \\  
   \hline 
  $\boldsymbol{\sigma}$ &  MPa  & 150 & 250 & 300 & 350 & 400 & 450 \\ 
  $\boldsymbol{\varepsilon}_{plastic}$ & -- & 0.00 & 0.06 & 0.30 & 1.00 & 2.50 & 5.00\\ 
\hline 
\end{tabular}
\label{table:1_copper_mat_properties_Elasto_Plastic}
\end{table}

\begin{figure}[hbt!]
    \centering
    \includegraphics[width=\linewidth]{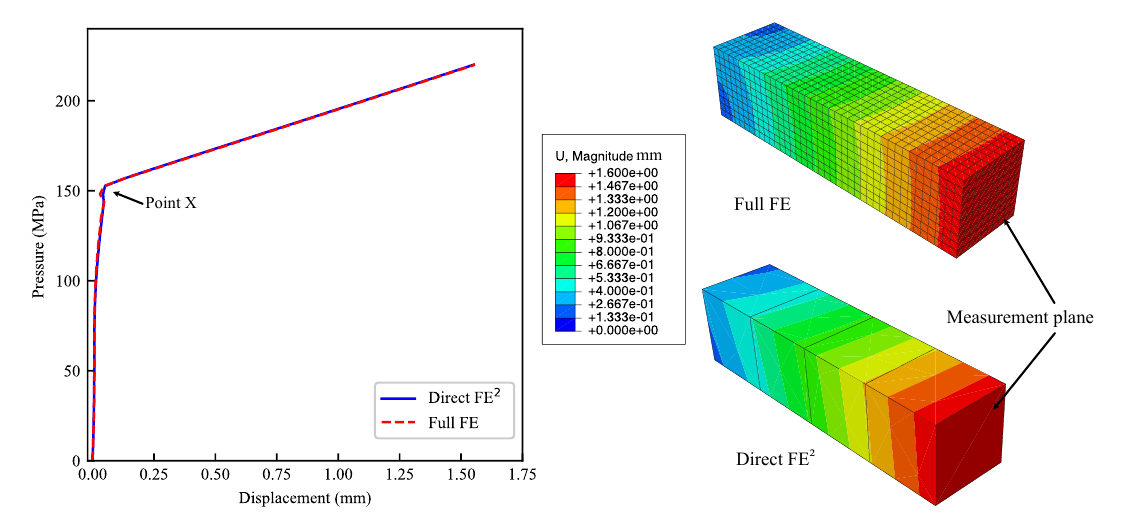}
    \caption{Comparison of pressure-displacement plots for Direct FE$^2$ and full FE approaches and displacement distribution contours in the last step of the simulation.}
    \label{fig:SG_Displacement}
\end{figure}

\begin{figure}[hbt!]
    \centering
    \includegraphics[width=\linewidth]{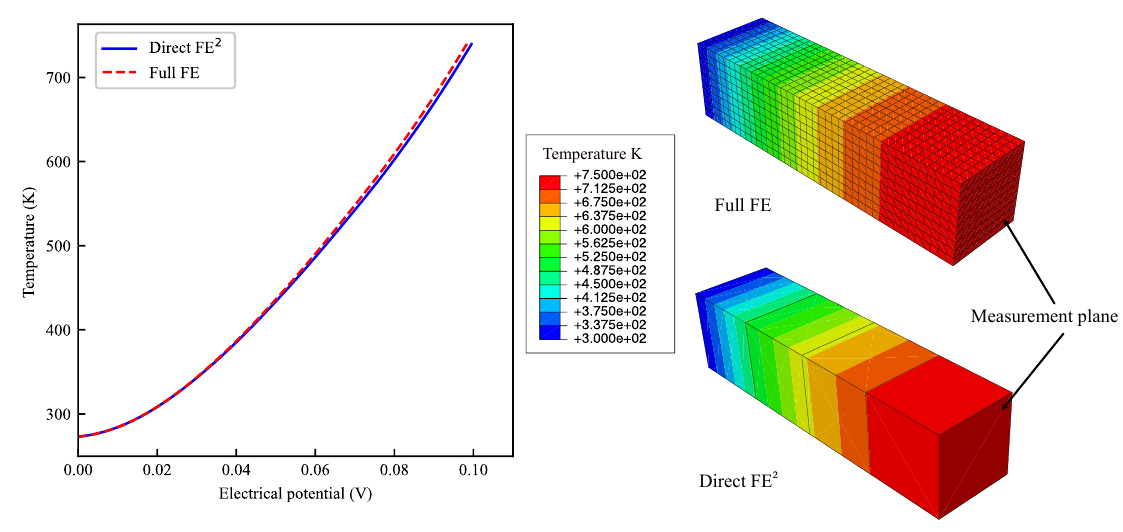}
    \caption{Comparison of temperature-electrical potential plots for Direct FE$^2$ and full FE approaches and temperature distribution contours in the last step of the simulation.}
    \label{fig:SG_Temperature}
\end{figure}

To compare the results of the Direct FE$^2$ and full FE models, displacement and pressure data are extracted on plane B. The pressure-displacement comparison is illustrated in Fig. \ref{fig:SG_Displacement}, showing almost an identical behavior from the two solutions. A slight discrepancy appears at point X, where a transition in the material model occurs. 
In addition, the displacement distribution at the end of the simulation is shown in Fig. \ref{fig:SG_Displacement}, which shows that the contours between the models are almost identical. Similarly, Fig. \ref{fig:SG_Temperature} compares temperature versus electrical potential plots, accompanied by temperature distribution visualizations. The results align well with the full FE approach, with maximum errors of 0.06\% for displacement and 0.02\% for temperature.

To further verify the Direct FE$^2$ approach, the impact of RVE mesh refinement on the simulation results is analyzed in the following sections. It is clear that the scaling factor in Eq. \eqref{eq:Scalling_factor} is directly linked to both the macro element size and the RVE size. An additional study is thus conducted, varying the macro element size while keeping the RVE size constant and adjusting the scaling accordingly. This will help us understand how the Direct FE$^2$ method responds to different macro mesh configurations, thereby facilitating the selection of an appropriate RVE or macro element size for various simulation scenarios.

\subsection{Effect of RVE mesh size at microscale}
To assess the impact of RVE mesh size on the accuracy of the proposed Direct FE$^2$ method, simulations were conducted using three different RVE mesh sizes: $\Delta$x = 0.1, 0.05, and 0.025 mm, while maintaining a constant macro element size of 10 mm. The results were compared to those of the full FE simulation, using mesh sizes similar to the RVEs. However, the full FE simulation showed no significant differences with mesh refinement. Therefore, for consistency in the comparison, a uniform mesh size of 1 mm was used.

\begin{figure}[hbt!]
    \centering
    \includegraphics[width=\linewidth]{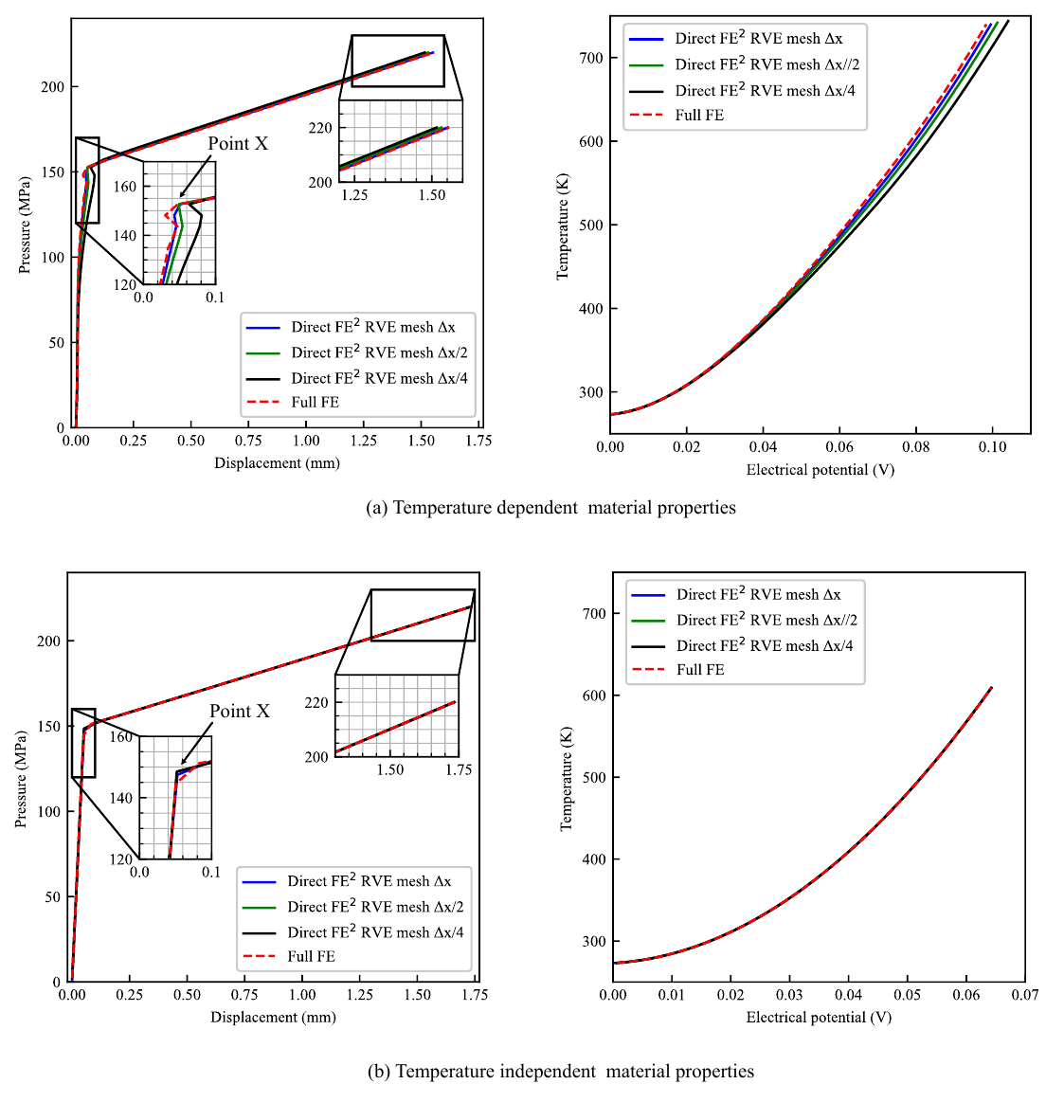}
    \caption{Effect of RVE mesh size on pressure-displacement and temperature-electric potential curves for temperature-dependent and independent thermoelectric material properties.}
    \label{fig:SG_RVE_mesh_size_Plots}
\end{figure}

\begin{figure}[hbt!]
    \centering
    \includegraphics[width=\linewidth]{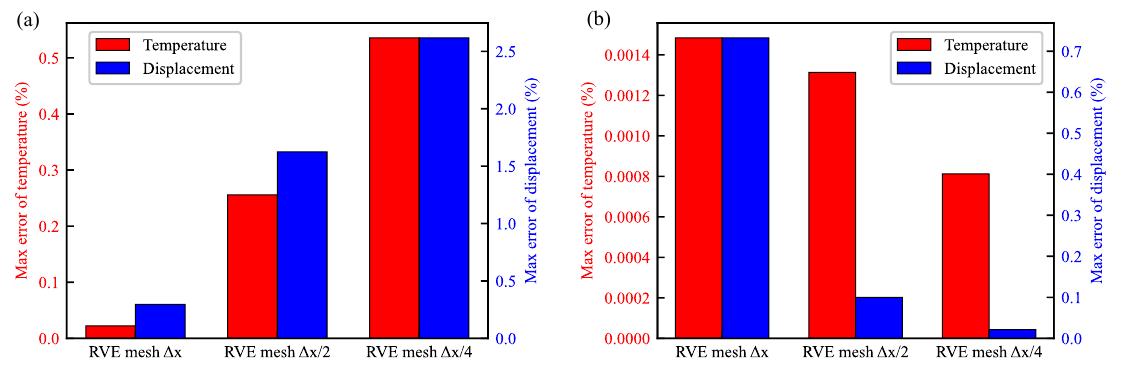}
    \caption{Effect of RVE mesh size on maximum errors of displacement and temperature for (a)  temperature-dependent and (b) temperature-independent thermoelectric material properties.}
    \label{fig:SG_RVE_mesh_size_Error}
\end{figure}

\begin{figure}[hbt!]
    \centering
    \includegraphics[width=\linewidth]{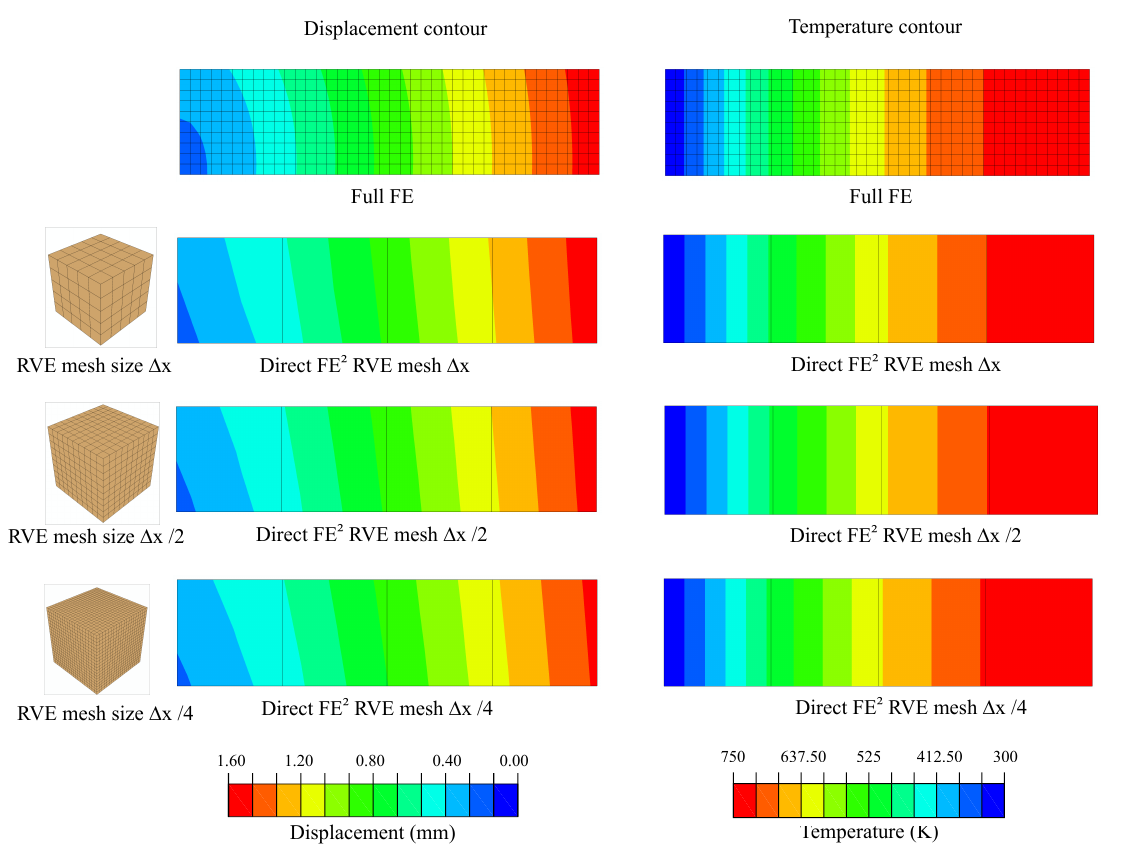}
    \caption{Effect of RVE mesh size on displacement and temperature contours.}
    \label{fig:2_2_Disp_Temp_Contours_RVE_All}
\end{figure}

As shown in Fig. \ref{fig:SG_RVE_mesh_size_Plots}a, the pressure-displacement curves for all three RVE mesh sizes closely match the full FE results. Nevertheless, a thorough analysis of the curves (inset images) indicates a discrepancy. Instead of converging towards the full FE result with mesh refinement, the results diverge. In particular, the maximum divergence at point X for the finer mesh is 2.85\%, in comparison to 0.575\% for the coarse mesh. A similar, more noticeable trend is seen in the temperature-electrical potential plots (Fig. \ref{fig:SG_RVE_mesh_size_Plots}a), where the finer RVE meshes show deviations from the full FE result with a maximum error of 0.5\%.
To further investigate this issue, additional simulations are performed using temperature-independent properties. As shown in Fig. \ref{fig:SG_RVE_mesh_size_Plots}b, these simulations show negligible differences in both displacement and temperature values when compared to the full FE result. The maximum error for displacement and temperature are 0.7\% and 0.002\%, respectively, for the coarse mesh.  In addition, Fig. \ref{fig:SG_RVE_mesh_size_Error} shows the maximum errors over the three RVE mesh sizes for both temperature-dependent and temperature-independent properties. It is evident that simulations performed with temperature-independent properties show better convergence to the full FE result compared to those with temperature-dependent properties. These investigations show that the discrepancy is due to nonlinear effects caused by the temperature-dependent properties of the RVEs and an overestimation of the temperature within the RVEs.

Fig. \ref{fig:2_2_Disp_Temp_Contours_RVE_All} illustrates the displacement and temperature distributions at the end of the simulation for all three RVE mesh sizes. These distributions closely align with the full FE results. Minor contour variations in displacement are observed at the constrained (fixed) point and the loading plane, which may be due to temperature-dependent properties. In contrast, no such variations are seen in the temperature contours.

\subsection{Effect of macro mesh size at macroscale}
\begin{figure}[hbt!]
    \centering
    \includegraphics[width=\linewidth]{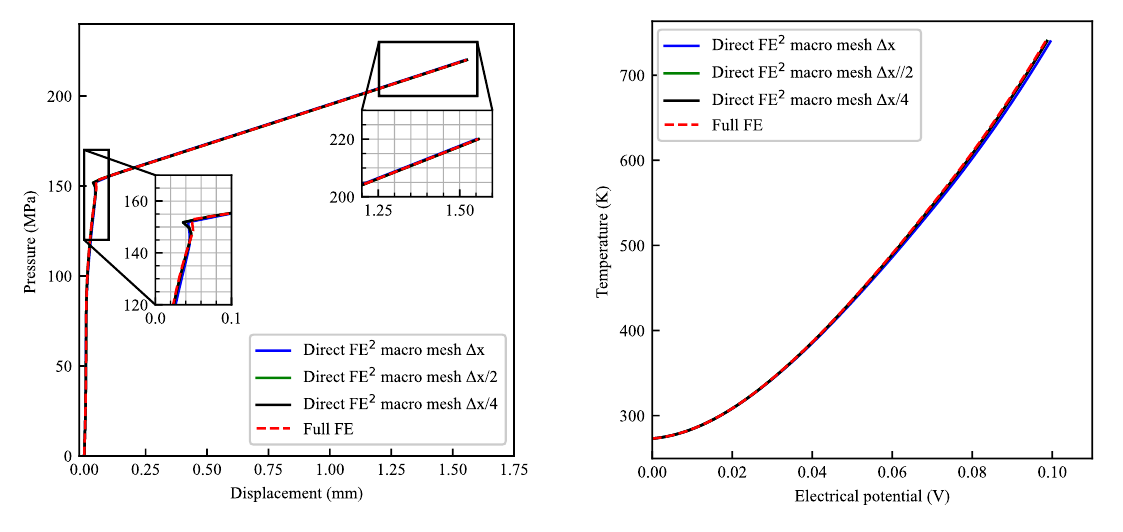}
    \caption{Effect of macro mesh size on pressure-displacement and temperature-electric potential curves. }
    \label{fig:SG_Macro_mesh_Disp_temp_plots}
\end{figure}

The impact of macro-scale mesh size on the accuracy of the Direct FE$^2$ method was assessed through a series of simulations using three different macro mesh sizes: $\Delta x$ = 10, 5, and 2.5 mm. This resulted in scaling factors for the RVEs of 5, 2.5, and 1.25, respectively. The RVE mesh size was kept constant at 0.1 mm across all simulations. The pressure-displacement results for the three macro mesh sizes are shown in Fig. \ref{fig:SG_Macro_mesh_Disp_temp_plots}, and minimal differences were observed when compared to the full FE results. Similarly, the temperature-electrical potential plots for all three macro mesh sizes show excellent alignment with the full FE simulation. This demonstrates a consistent and close agreement across various levels of macro mesh refinement.

\begin{figure}[hbt!]
    \centering
    \includegraphics[width=0.5\linewidth]{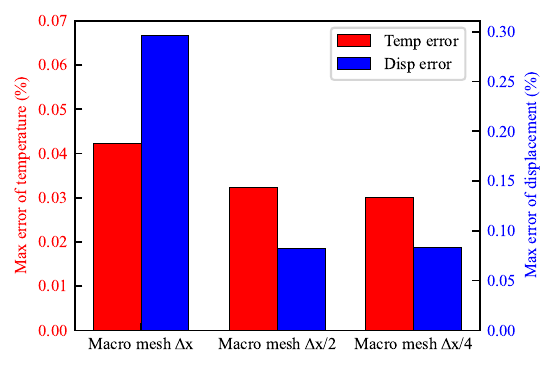}
    \caption{Effect of macro mesh size on maximum errors of displacement and temperature.}
    \label{fig:SG_macro_mesh_Error}
\end{figure}

Additionally, it can be seen that further refinement of the macro mesh size diminishes the influence of the RVEs, causing the model to behave more like a full FE model. Fig. \ref{fig:SG_macro_mesh_Error} illustrates the maximum errors for both temperature and displacement across the different macro mesh sizes. These results show that refining the macro mesh element size significantly reduces the maximum error relative to the full FE results. This finding is consistent with theoretical expectations, as a finer macro mesh typically improves accuracy by closely approximating the full FE model.

\begin{figure}[hbt!]
    \centering
    \includegraphics[width=\linewidth]{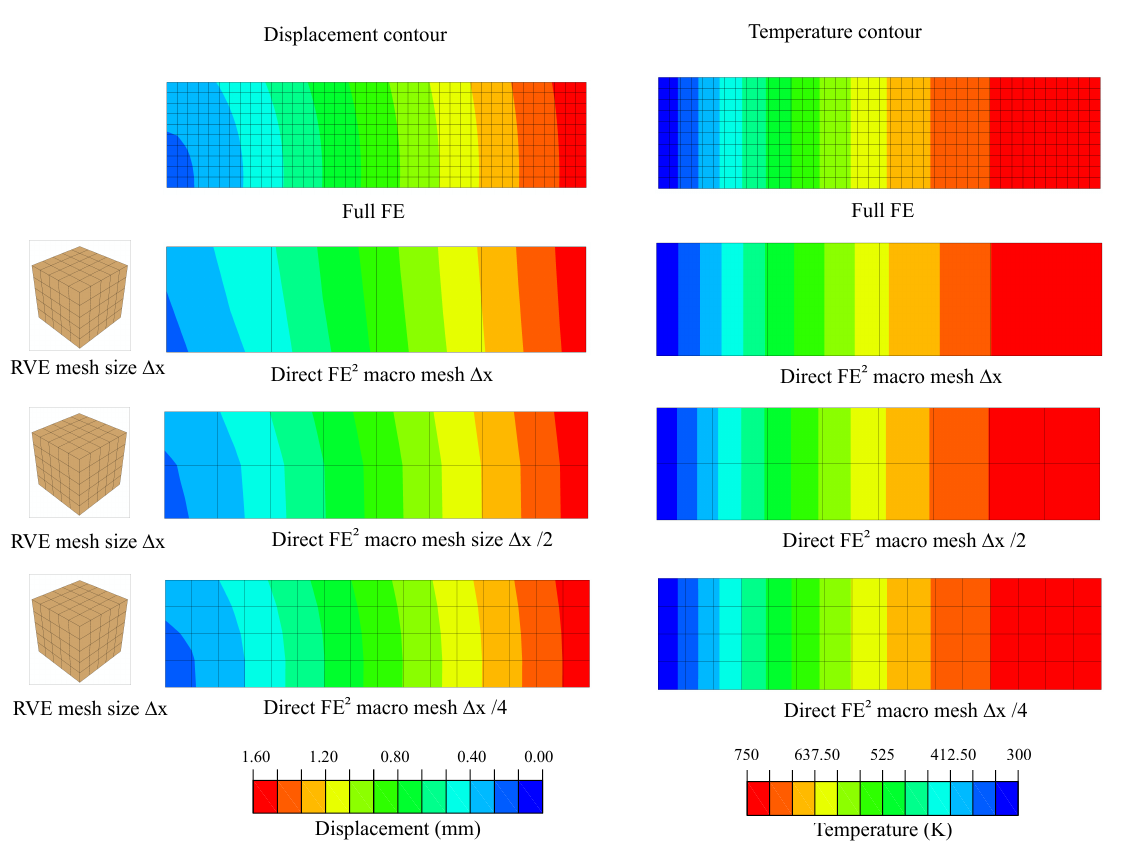}
    \caption{Effect of macro mesh size on displacement and temperature contours.}
    \label{fig:SG_Macro_mesh_Temp_Disp_plots}
\end{figure}

The predicted displacement and temperature distributions at the end of the simulation for all three macro mesh sizes are presented in Fig. \ref{fig:SG_Macro_mesh_Temp_Disp_plots}. These distributions align well with the full FE results. However, at the finest macro mesh size ($\Delta x/4$), the displacement and temperature distributions show an even closer match. Based on these findings, it can be concluded that the scaling factor has only a marginal impact on the accuracy of the Direct FE$^2$ method. However, if the macro element size is too large, accurately capturing gradient changes in kinematic variables under complex constraint conditions becomes difficult. This effect of the scaling factor (macro mesh size) is consistent with observations made by Cui et al. \cite{cui2024direct} in purely mechanical systems.
\section{Application to SPS}
The proposed Direct FE$^2$-based multiscale modeling framework is applied to powder compaction and sintering simulations. First, we simulate the purely mechanical compaction of powder particles and compare the results with those from the full FE method to validate the applicability of the Direct FE$^2$ approach. Next, we extend the method to a coupled electro-thermal-mechanical analysis, examining interactions within the powder compaction and sintering process. This coupled Direct FE$^2$ simulation is then compared to a coupled full FE simulation, focusing on computational efficiency and predictive accuracy.

For comparison purposes, we begin by considering a small segment of powder packing arranged in a simple cubic geometry. Following this, we apply the coupled Direct FE$^2$ method to different RVE geometries, comparing their densification behavior and validating the method by comparing the results to the Heckel equation, which describes the relationship between applied pressure and powder densification during compaction. Lastly, we simulate a larger portion of the SPS-ed sample under two different current inputs, and the results are discussed in detail.

\subsection{Powder compaction: mechanical simulation}
\label{sec:powder_compaction}
\begin{figure}
    \centering
    \includegraphics[width=0.75\linewidth]{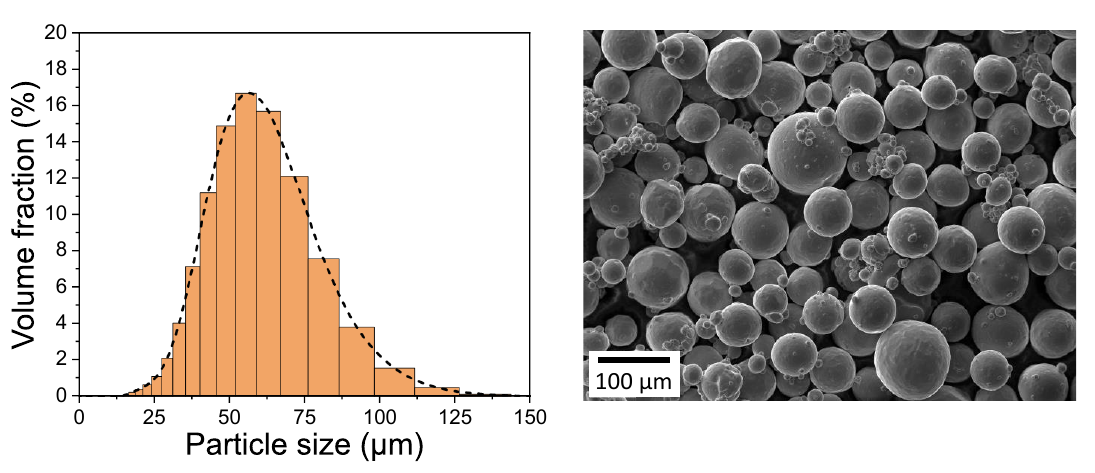}
    \caption{Particle size distribution and scanning electron microscope (SEM) image of the copper powder.}
    \label{fig:Powder_exp}
\end{figure}

\begin{figure}[hbt!]
    \centering
    \includegraphics[width=0.75\linewidth]{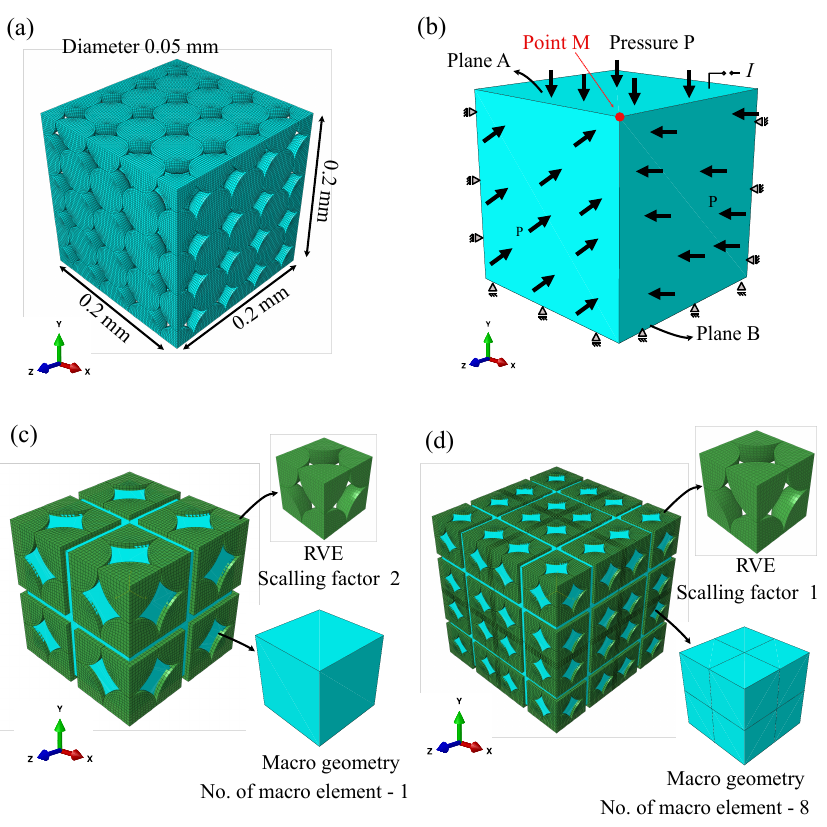}
    \caption{(a) Geometric representation for powder compaction simulation in a full FE framework, (b) boundary conditions along with Direct FE$^2$ models with (c) a single macro element and (d) eight macro elements.}
    \label{fig:PG_Geometry}
\end{figure}

A cubic section of powder packing with dimensions of $0.2\times0.2\times0.2$ mm is selected for the mechanical compaction analysis, as shown in Fig. \ref{fig:PG_Geometry}a. The particles are assumed to be spherical with a uniform diameter of 50 $\mu$m. The particle size distribution and micrograph obtained from the experiment (Fig. \ref{fig:Powder_exp}) confirm that this assumption is reasonable for the copper powder used in the numerical simulations in this study. A uniform compaction boundary condition, commonly used in powder compaction analysis, is applied. Specifically, three sides of the geometry are constrained in the perpendicular direction, while the remaining three surfaces are subjected to a pressure of 220 MPa, as illustrated in Fig. \ref{fig:PG_Geometry}b. The corresponding Direct FE$^2$ model of the same size is composed of RVEs with dimensions of $0.05\times0.05\times0.05$ mm, as shown in Fig. \ref{fig:PG_Geometry}c. In this model, the macro geometry is represented by a single macro element, resulting in a scaling factor of 2 for the RVE. To explore the impact of the scaling factor on powder geometry, an additional Direct FE$^2$ model with 8 macro elements is considered, keeping the RVE size identical. This configuration results in a scaling factor of 1, as depicted in Fig. \ref{fig:PG_Geometry}d. A mesh size of 5 $\mu$m is used for both the full FE model and the RVEs in the Direct FE$^2$ models.

The electro-thermal material properties used in this study are summarized in Table \ref{table:1_copper_mat_properties_ET}. Temperature-independent properties are assumed at a constant temperature of 300 K. Additionally, an elastoplastic material model, detailed in Table \ref{table:1_copper_mat_properties_Elasto_Plastic}, is used for comparison. The material has a Young's modulus of 115 GPa and a Poisson’s ratio of 0.34. The material properties are assigned to both the full FE model and the RVEs in the Direct FE$^2$ models, while dummy material properties are assigned to the macro elements in the Direct FE$^2$ models. A friction coefficient of 0.3 is assumed for particle contacts, with hard contact enforced for normal interactions, allowing no separation after contact. Contact stabilization is applied for the full FE model.

\begin{figure}[hbt!]
    \centering
    \includegraphics[width=0.6\linewidth]{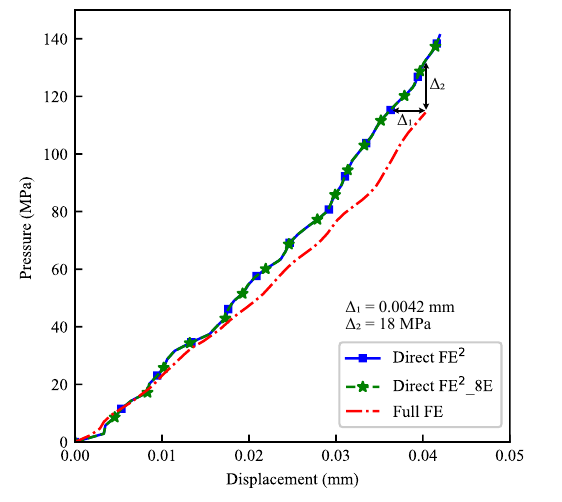}
    \caption{Comparison of pressure-displacement curves obtained from full FE and Direct FE$^2$ simulations in the mechanical powder compaction test.}    
    \label{fig:PG_Disp_Mech}
\end{figure}

\begin{figure}[hbt!]
    \centering
    \includegraphics[width=0.75\linewidth]{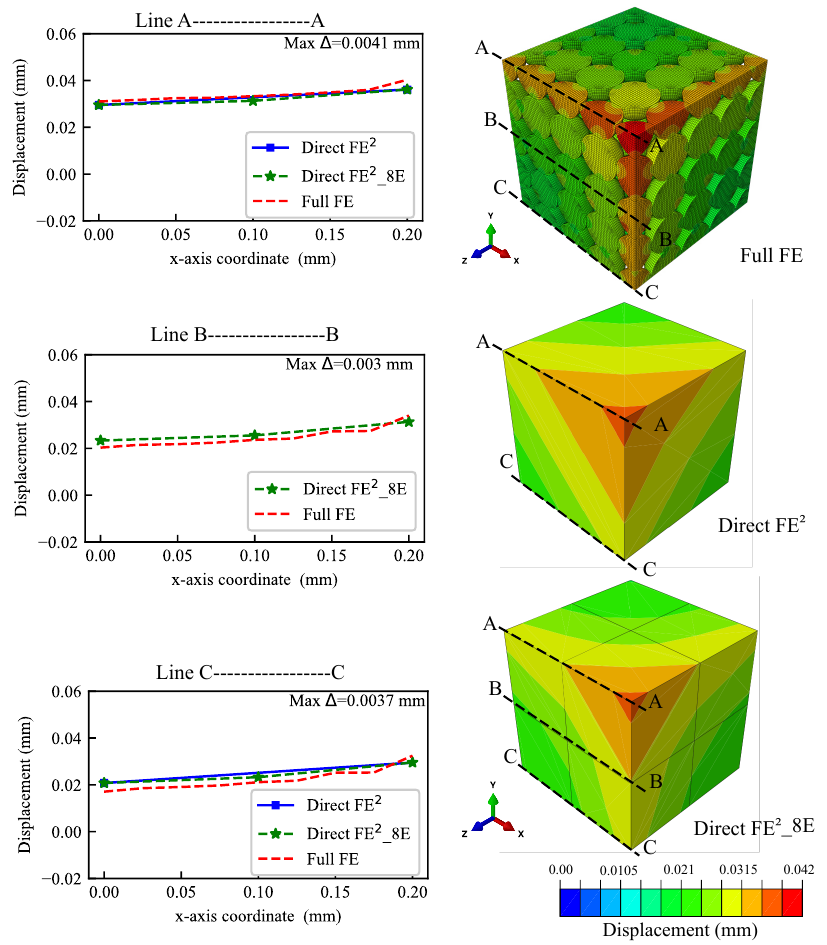}
    \caption{Comparison of displacement distribution for FE and Direct FE$^2$ simulations in the mechanical powder compaction test.}
    \label{fig:PG_Disp_Contour_Mech}
\end{figure}

To compare the results, displacement values are extracted at Point M (maximum deflection point) - see the red corner indicated in Fig. \ref{fig:PG_Geometry}b. A comparison of the pressure-displacement relationship between the full FE method and the Direct FE$^2$ method (with one and eight macro elements, respectively) is presented in Fig. \ref{fig:PG_Disp_Mech}. The results confirm a very similar trend from the two approaches during the initial stage, which is characterized by low compaction. However, at a later stage, when compaction increases, a deviation of 0.0042 mm at a pressure of 115 MPa is observed. This deviation can be attributed to the cumulative effects of increased damping or instability in the full FE model, which is a consequence of the larger number of contacts and particle arrangements in the full FE geometry.
Similarly, this observation can be attributed to the fact that the full FE model exhibits nearly the same displacement (0.04 mm) at 18 MPa lower pressure than the Direct FE$^2$ model. The pressure-displacement curves from the Direct FE$^2$ model with eight macro elements (scaling factor of one) closely align with the model using one macro element, indicating that the effect of the scaling factor on the displacement magnitude at point M is negligible. 

Fig. \ref{fig:PG_Disp_Contour_Mech} shows the displacement distribution for the full FE and the Direct FE$^2$ models under a pressure value of 115 MPa. The distributions across all three cases appear identical. However, a detailed assessment of the displacement values along the A-A, B-B, and C-C axes reveals maximum deviations of 0.0041 mm, 0.0030 mm, and 0.0037 mm, respectively. The maximum deviation occurs along the A-A axis at x-coordinate 0.2 mm (point M), which experiences the combined influence of all three pressure components, resulting in the highest deflection. In contrast, the deviations along the B-B and C-C axes are relatively small due to the smaller displacement magnitudes experienced along these axes. The geometry is presented in its undeformed configuration to provide a clearer comparison of the displacement distribution.

\subsection{Powder sintering: electro-thermal-mechanical simulation}
The coupling of electro-thermal-mechanical effects introduces significant nonlinearity into the model. Additionally, the inclusion of contact interactions leads to convergence issues in the FE model. To enable a comparison between the Direct FE$^2$ model and the full FE model results, we apply two simplifications to the model: 1) Linear elastic material model; and 2) Temperature-independent thermo-electric material properties.

The same geometry from the previous section is used, with a mesh generated using coupled elements (Q3D8). Similar contact properties as in the previous section are applied, and the electrical and thermal resistances between particle surfaces are assumed to be negligible when in contact. The mechanical boundary conditions are also kept unchanged. Additionally, a surface current is applied to one end of the geometry, while the opposite end is grounded. To create a temperature gradient, a constant temperature of 300 K is imposed on plane A, as illustrated in Fig. \ref{fig:PG_Geometry}b. To evaluate the effect of RVE scaling on the coupled simulation, two Direct FE$^2$ cases are considered: one with a single macro element and the other with eight macro elements.

\begin{figure}
    \centering
    \includegraphics[width=\linewidth]{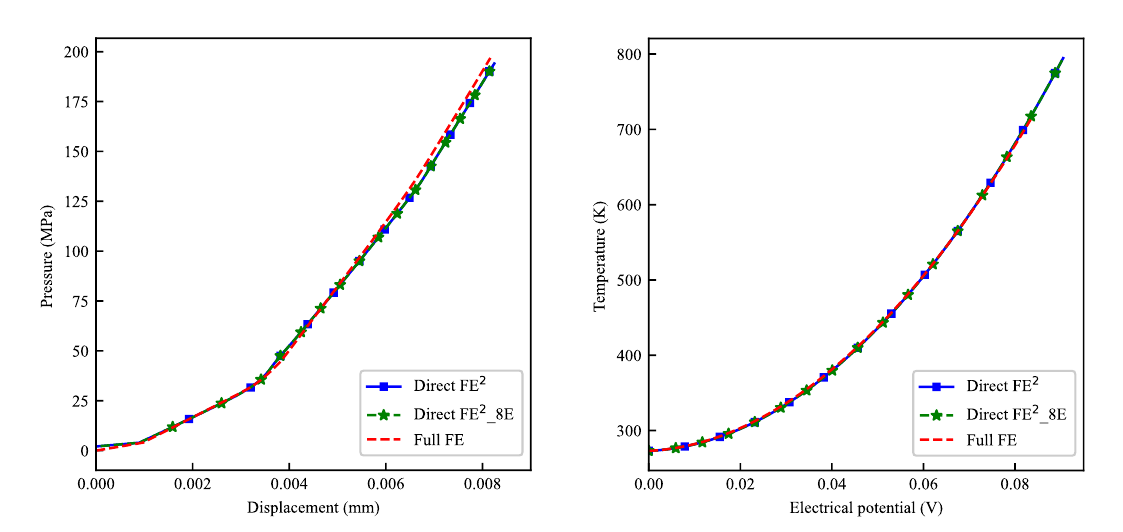}
    \caption{Comparison of pressure-displacement and temperature-electrical potential curves obtained from full FE and Direct FE$^2$ simulations for fully coupled powder sintering.}
    \label{fig:PG_ETM_Plots}
\end{figure}

Displacement and temperature values are again extracted at Point M (Fig. \ref{fig:PG_Geometry}b) for comparison. These values are plotted in Fig. \ref{fig:PG_ETM_Plots} to illustrate the pressure-displacement and temperature-electrical potential relationships. Although the pressure-displacement plots for the Direct FE$^2$
and full FE methods overlap to a large extent, small deviations are observed at the end of the simulation, which remain below 1\%. However, the pressure-displacement plots reveal that the full FE method provides better alignment in the coupled case compared to the pure mechanical case (as discussed in Section \ref{sec:powder_compaction}). It is important to note that the material model used here is elastic, resulting in less compaction compared to the elastoplastic material model used in the pure mechanical case. Similarly, the temperature-electric potential results show identical trends. Furthermore, in the coupled case, the effect of the scaling factor is minimal, in agreement with the results of the Section \ref{sec:powder_compaction}.

\begin{figure}
    \centering
    \includegraphics[width=0.75\linewidth]{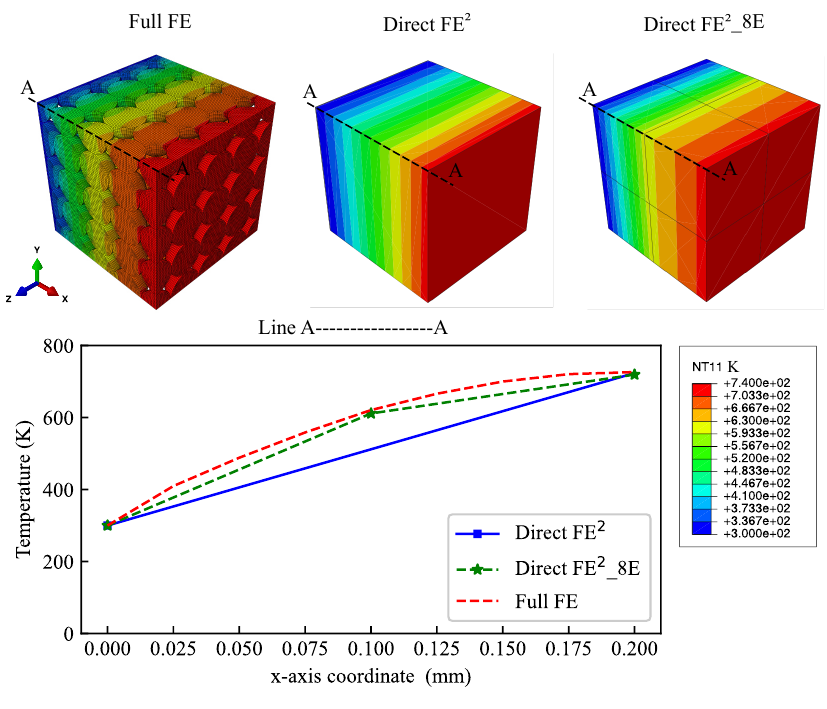}
    \caption{Temperature distribution from full FE and Direct FE$^2$ simulations for fully coupled powder sintering.}
    \label{fig:PG_Temp_Disp_ETM}
\end{figure}

Fig. \ref{fig:PG_Temp_Disp_ETM} shows the temperature distributions at a fixed electrical potential within the simulation for both the full FE and Direct FE$^2$ solutions. The temperature data are plotted at the element nodes along axis A-A for all three cases. In the Direct  FE$^2$ model, using one and eight macro elements, the temperature values at the measured node demonstrate maximum deviations of only 5 K (at x-coordinate 0.2 mm) and 9 K (at x-coordinate 0.1 mm), respectively. In the Direct FE$^2$ model, linear elements are employed for the simulation, resulting in the linear interpolation of the temperature distribution. Consequently, the deviation at the midpoint of the elements is comparatively higher. For instance, in the case of a single macro element at the x-coordinate of 0.1 mm, the deviation reaches approximately 90 K in comparison to the full FE model.
Increasing the number of macro elements improves the approximation accuracy for these variables. Note that the degrees of freedom (i.e., displacement, temperature, and electrical potential) in the Direct FE$^2$ model are comparable to those in the full FE model, but only at the macro level. For further analysis, only one macro element is considered for the Direct FE$^2$ method.

\subsubsection{Computational performance}
In the Direct FE$^2$ method, the number of degrees of freedom (DOF) can be reduced by increasing the scaling factor, which enables calculation savings and runtime acceleration. To demonstrate the computational efficiency of the proposed Direct FE$^2$ method, a comparative analysis is conducted using three different mesh sizes that affect the DOF of the model. The Direct FE$^2$ model is selected with a macro element that has a scaling factor of 2. All simulations are performed on an AMD Ryzen 7 3700X 8-Core Processor and an NVIDIA Quadro RTX 4000, utilizing Abaqus's parallel processing capabilities with both CPU and GPU.

\begin{figure}[hbt!]
    \centering
    \includegraphics[width=1\linewidth]{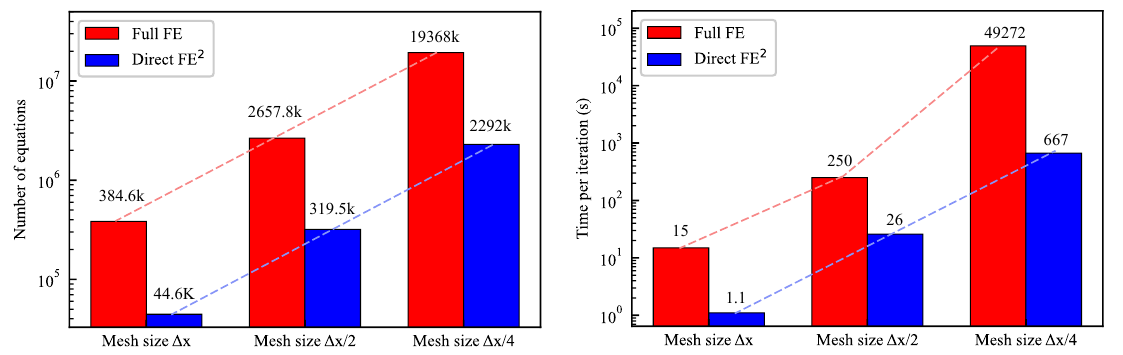}
    \caption{Comparison of the number of equations and the time required per iteration for the full FE and the Direct FE$^2$ for the fully coupled powder sintering.}
    \label{fig:PG_Time_Equation}
\end{figure}

A comparative analysis of the number of equations and the computational time per iteration is computed and shown in Fig. \ref{fig:PG_Time_Equation} for the full FE and Direct FE$^2$ approaches. Reducing the mesh size increases the number of degrees of freedom, resulting in a corresponding increase in the number of equations and computation time. However, as expected, the Direct FE$^2$ model demonstrates a significant reduction in both the number of equations and computational time compared to the full FE model. Specifically, for a fixed mesh size, the number of equations (NOE) in the Direct FE$^2$ model is approximately eight times lower, which corresponds to a volumetric scaling factor of eight (resulting from a scaling factor of two in each spatial direction). Further increases in the scaling factor can lead to even greater reductions in both the number of equations and associated computational time.

Additionally, halving the mesh size results in an eight-fold increase in the number of equations. While both methods exhibit this trend, the full FE model experiences a more significant increase in computational time compared to the Direct FE$^2$ method. For instance, when the mesh size is reduced from $\Delta x$ to $\Delta x/2$, the Direct FE$^2$ method operates approximately nine times faster than the full FE model. When the mesh size is further reduced to $\Delta x/4$, this speed advantage increases to a factor of 70. These comparisons highlight the considerable computational efficiency of the Direct FE$^2$ method, particularly regarding reduced computational time and resource usage. These performance reports and speedup results highlight the applicability and robustness of the method for simulations involving coupled electrical, thermal, and mechanical phenomena in larger geometries.

\subsection{Powder morphology}
\label{sec:morphology}
\begin{figure} [hbt!]
    \centering
     \includegraphics[width=1\linewidth]{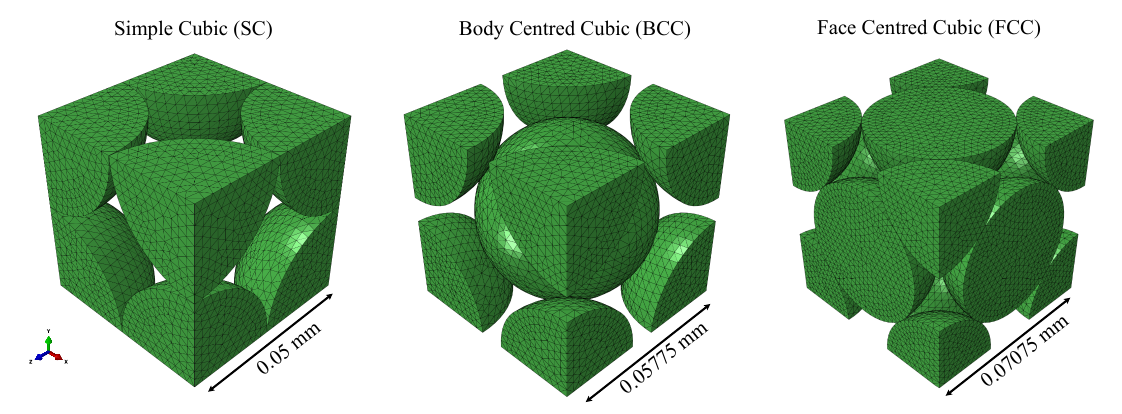}
    \caption{RVE geometries employed for the present powder morphology analyses.}
    \label{fig:RVE_Geo_All}
\end{figure}

This section explores three different types of RVEs in order to accommodate more realistic geometries of powder arrangement beyond a simple cubic pattern, including simple cubic (SC), body-centered cubic (BCC), and face-centered cubic (FCC) - seeFig. \ref{fig:RVE_Geo_All}. The initial packing densities for SC, BCC, and FCC structures are 52.36\%, 68.01\%, and 74.05\%, respectively. The geometric dimensions of each of the three RVE types (SC, BCC, and FCC) are 0.05, 0.05775, and 0.07075 mm, respectively. These dimensions are scaled in the Direct FE$^2$ model according to the scaling factor defined in Eq. \eqref{eq:Scalling_factor}. To facilitate a mesh-independent comparison, a fixed mesh size of \(0.05 \times \Bar{\alpha}_{\xi}\) is employed for all three RVEs.

In this comparison, the Johnson-Cook material model is employed for copper, a widely accepted model in the context of hot isostatic pressing applications \cite{peng2021compaction}. The material parameters used in this study for copper are taken from \cite{dai2011numerical}. For more details concerning the constitutive model, see \ref{app:App_JC_model}. These RVEs are analyzed under two different boundary conditions: uniform compression and uniaxial compression. Here, uniaxial compression is used to represent SPS loading conditions, wherein the load is applied in a uniaxial direction. In both scenarios, an electric current is applied at Plane A, while Plane B is maintained at ground voltage (V = 0 V).

\begin{figure}
    \centering
    \includegraphics[width=1\linewidth]{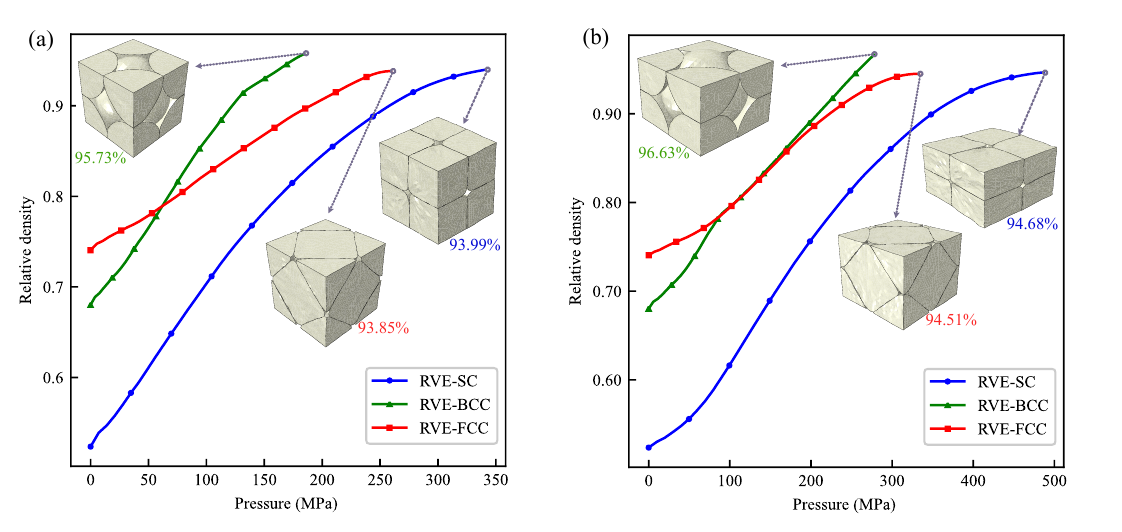}
    \caption{Relative density versus pressure curves for three types of RVE under different boundary conditions: (a) uniform compression, (b) uniaxial compression.}
    \label{fig:RVEs_All_RD_Plots}
\end{figure}

A key measure in evaluating the powder compaction process involved in SPS is relative density. Fig. \ref{fig:RVEs_All_RD_Plots} illustrates how relative density changes with applied pressure for the three different RVEs: SC, BCC, and FCC, under two different compression scenarios. The deformed shapes of the three RVEs are also depicted in these plots for clearer interoperability.

The densification rate varies considerably depending on the powder morphology. Furthermore, the application of pressure under different conditions also influences the compaction rate. In the cases of SC and FCC structures, the densification rate decreases at higher densities across both boundary conditions. In contrast, the behavior of the BCC configuration is anomalous, exhibiting a slight kink in its relative density curve. This kink arises from high compression, which misaligns the central particle in the BCC structure, thereby disrupting the smooth compression process.

This anomaly is associated with the application of boundary conditions. Under periodic boundary conditions, the constraint nodes are situated exclusively on the external surfaces of the RVEs, allowing the central particle to move beyond the boundaries of the RVE, unrestricted by external constraints. However, the central nodes of particles in a BCC structure are linked to the macro elements, as outlined in the relevant equations. This phenomenon is particularly evident under conditions of uniform compression at high pressures (Fig. \ref{fig:RVEs_All_RD_Plots}a) and uniaxial compression at lower pressures (Fig. \ref{fig:RVEs_All_RD_Plots}b). The deformed shape of the BCC RVE further illustrates this behavior, showing greater compression and relative density compared to the SC and FCC unit cells.

Moreover, the FCC structure, which exhibits a higher initial density, requires less pressure to achieve a specific density than the BCC structures. These findings underscore the importance of fully constraining any free particles within the RVE to prevent such anomalies in the results and highlight the significant impact of the initial particle arrangement on compaction efficiency under pressure.

\subsection{Validation}
\label{sec:validation}
To validate the present Direct FE$^2$ framework, simulation results are compared with the analytical solution of the Heckel equation, which describes the decrease in porosity of a powder under applied pressure during compaction as \cite{tavakoli2005study, denny2002compaction}:
\begin{equation}
     \ln[1 / (1 - \rho_d)] = mP + c
\end{equation}

where $\rho_d$ is the relative density, $P$ represents applied pressure, and $c$ and $m$ are constants. The constant $m$, known as the densification coefficient, indicates that materials with a higher $m$ value can achieve a higher relative density at a constant pressure.

\begin{figure}
    \centering
    \includegraphics[width=1\linewidth]{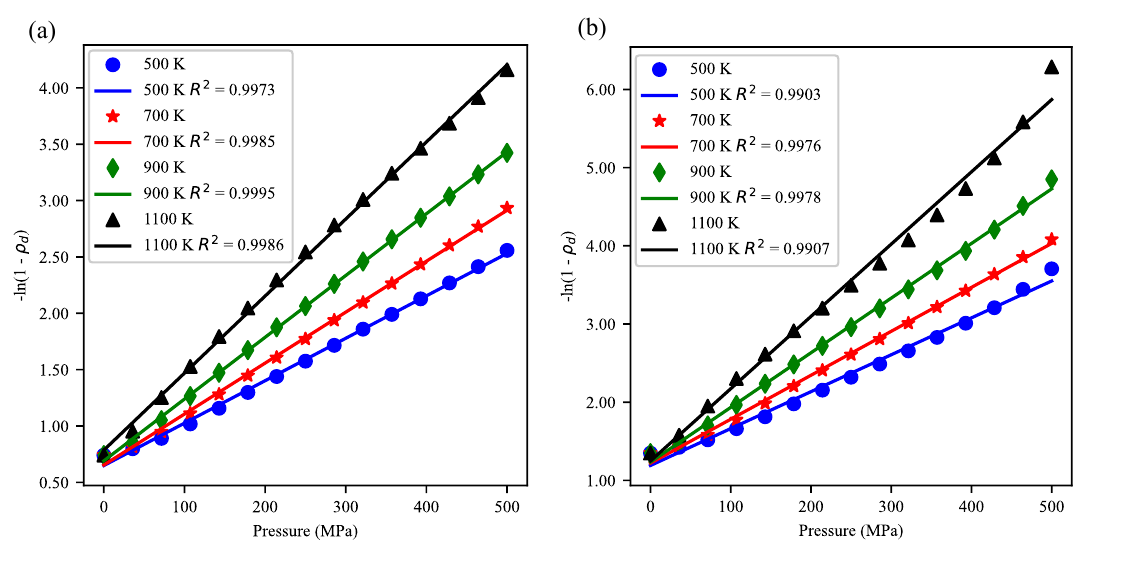}
    \caption{Fitting of the Direct FE$^2$ simulation results to the Heckel equation for uniaxial compaction of (a) SCC and (b) FCC powder morphologies.}
    \label{fig:Heckel_eq_plots}
\end{figure}

The Direct FE$^2$ model uses a configuration similar to that described in Section \ref{sec:morphology}, along with the same material model. Simulations are conducted at constant temperatures of 500, 700, 900, and 1100 K for SCC and FCC RVE morphologies under uniaxial compression. As shown in Fig. \ref{fig:Heckel_eq_plots}, the simulation results align closely with the Heckel equation, with an $R^2$ value exceeding 0.99. Furthermore, at higher temperatures, the slope $m$ of the fitted line increases, suggesting that densification becomes easier as requires less pressure. 

\subsection{Multiscale sintering simulation}
\begin{figure}
    \centering
    \includegraphics[width=1\linewidth]{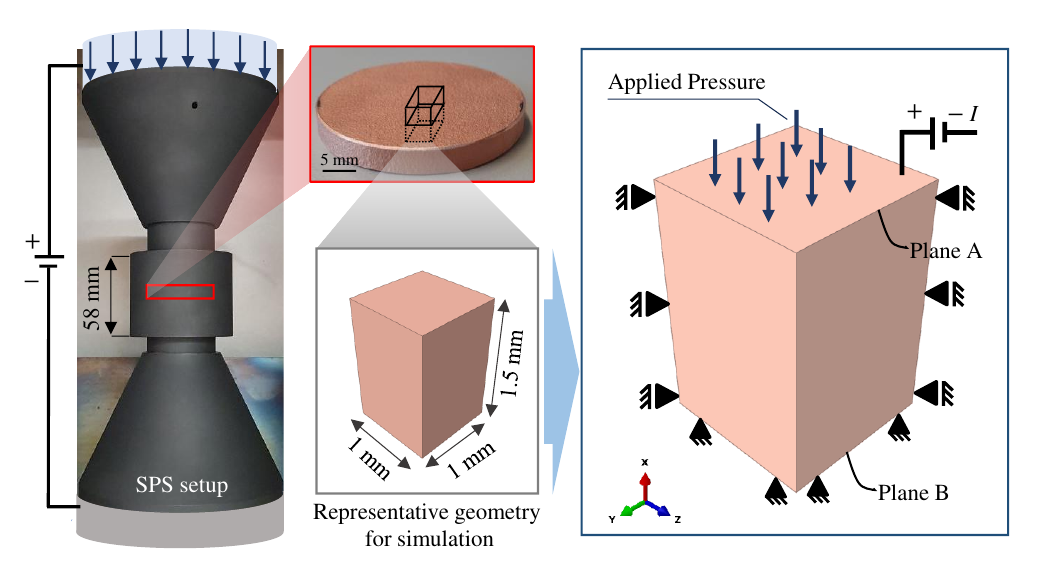}
    \caption{Schematic representation of the sample geometry under consideration for a coupled SPS process analysis and the boundary conditions used in the analysis.}
    \label{fig:SPS_main_Geo}
\end{figure}

The representative geometry chosen for the final simulation, as illustrated in Fig. \ref{fig:SPS_main_Geo}, has dimensions of $1\times1\times1.5$ mm. During the SPS process, both load and current are applied at one end, while the opposite end is fixed and grounded (V = 0 V), establishing a uniaxial boundary condition. In this setup, an electric current is introduced at plane A, with plane B grounded at 0 V. All side surfaces are constrained against movement perpendicular to their respective planes. To prevent any rigid body motion, a single point on plane B is fixed. Two scenarios are analyzed, each with different current inputs that affect the heating rate.

\begin{figure}
    \centering
    \includegraphics[width=0.75\linewidth]{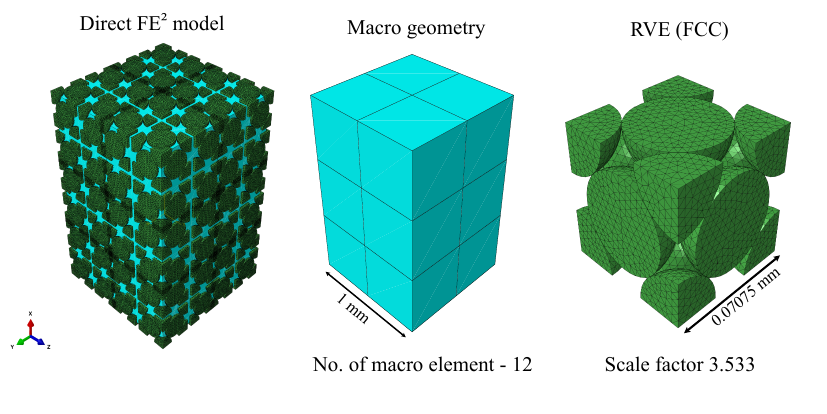}
    \caption{Direct FE$^2$ model of the geometry with macro elements and a RVE of type FCC.}
    \label{fig:SPS_Geo}
\end{figure}

In this test, the RVE type is assumed to be FCC, which is more representative of powder packing. The macro geometry is discretized into 12 elements, resulting in a scaling factor of 3.533. The Direct FE$^2$ model for the given geometry, incorporating both the macro elements and the RVE, is shown in Fig. \ref{fig:SPS_Geo}. The Johnson-Cook material model is used, as described in Section \ref{sec:morphology}.

\begin{figure}
    \centering
    \includegraphics[width=1\linewidth]{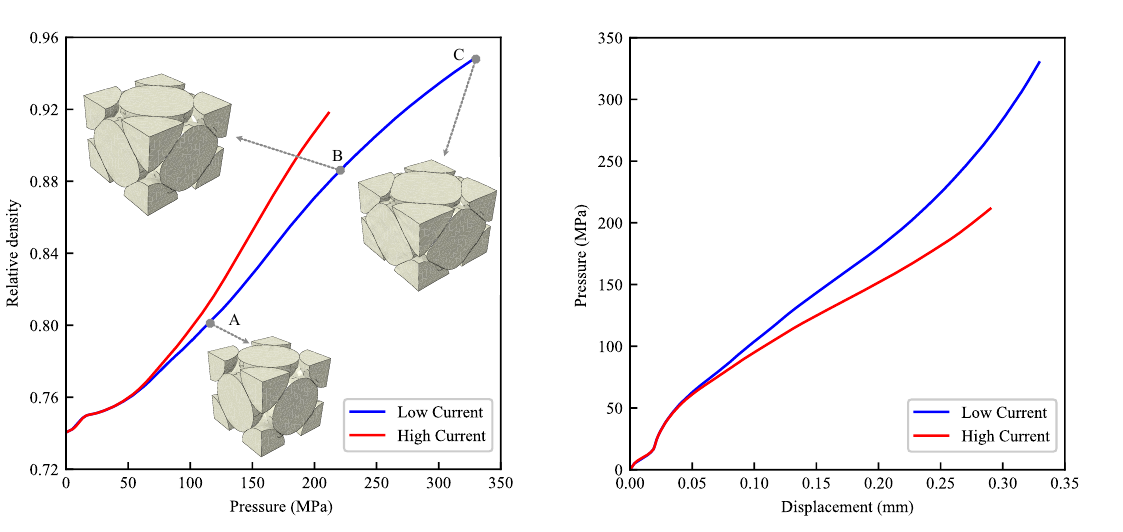}
    \includegraphics[width=1\linewidth]{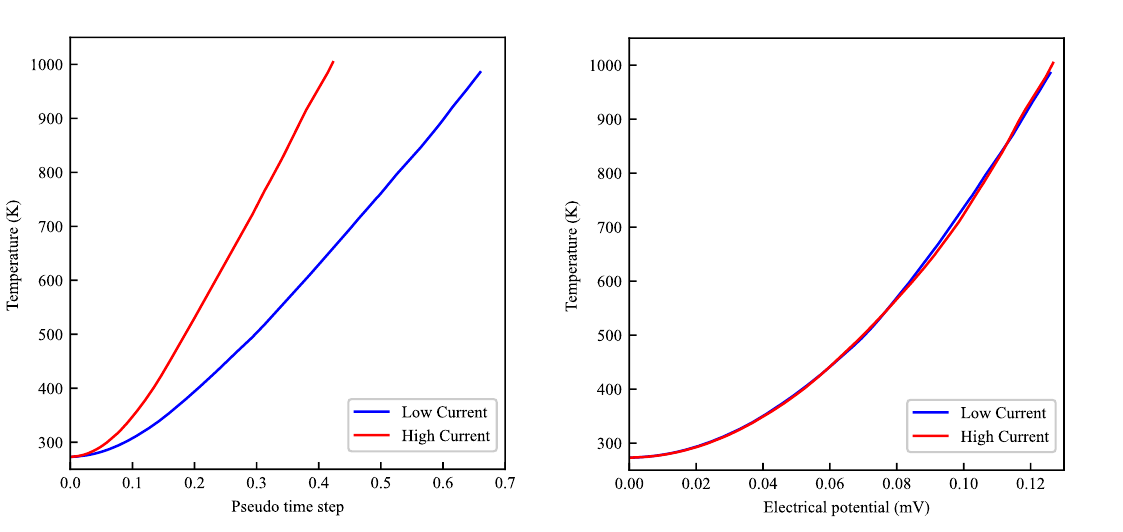}
    \caption{Relative density vs. applied pressure, pressure vs. displacement, temperature vs. pseudo time step and temperature vs. electrical potential curves obtained for the two surface current applications on the SPS geometry.}
    \label{fig:SPS_Pressure_plots}
\end{figure}

Fig. \ref{fig:SPS_Pressure_plots} shows the plots of relative density versus applied pressure and pressure versus displacement for two different current inputs. It is clear that the densification rate increases with higher current input and decreases with lower current input. However, in the SPS process, lower pressures are typically used due to the machine’s safety limit, which ranges from 20 to 100 MPa. We chose a higher pressure to accelerate the densification process and reduce computational time. Both graphs demonstrate that, for a given relative density or displacement, higher pressure is required in the low-current scenario. The inset images display the deformed configuration of the middle RVE geometry at relative densities of 0.80 (A), 0.88 (B), and 0.95 (C). At the start of the process, a sudden change is observed due to the initial contact between all particles. In the Direct FE$^2$ model, the initial geometry maintains a slight gap (in microns) between particles, requiring a small force to establish contact. From the temperature versus pseudo time-step plots, a clear change in heating rate is observed for higher and lower current inputs.

However, the temperature versus electrical potential plots appear similar due to the consistent geometric configuration in both cases. Therefore, the temperature increase is solely attributed to Joule heating. Notably, at higher current inputs, the maximum density is observed to be lower than at lower current inputs. This is due to the increased nonlinearity (both geometrical and material) in the solver. However, by carefully considering solution strategies such as adaptive mesh refinement and stabilization techniques, it is possible to achieve higher compression.

\begin{figure}
    \centering
    \includegraphics[width=0.75\linewidth]{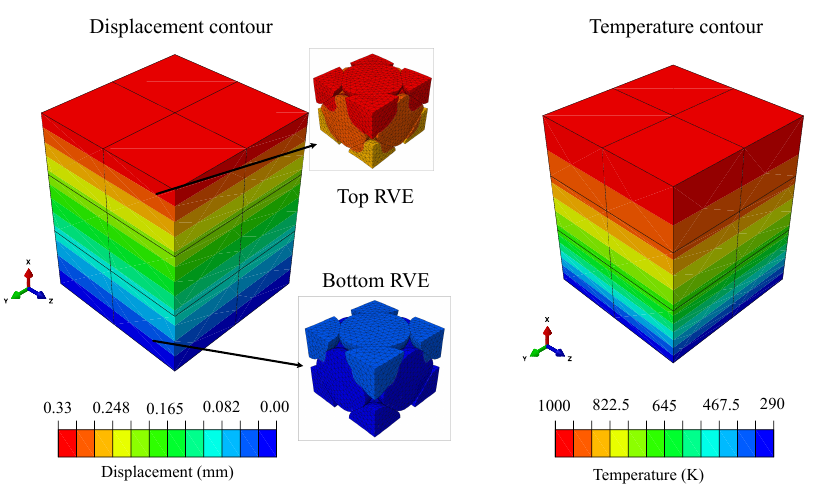}
    \caption{Displacement and temperature distribution from the fully coupled SPS simulation at lower current input.}
    \label{fig:SPS_disp_temp}
\end{figure}

Fig. \ref{fig:SPS_disp_temp} illustrates the displacement and temperature distribution at the end of the simulation for the low-current input case. The displacement diagrams depict the deformed shapes of the RVEs from both the top and bottom macro elements. It is clear that the top RVE undergoes greater compression than the bottom RVE, indicating uneven compaction. This is due to the temperature gradient present in the geometry. In the SPS process, temperature gradients play a significant role in density variation. Additionally, the literature suggests that sintered samples exhibit slight density gradients, if any, due to these temperature gradients.

\begin{figure}
    \centering
    \includegraphics[width=\linewidth]{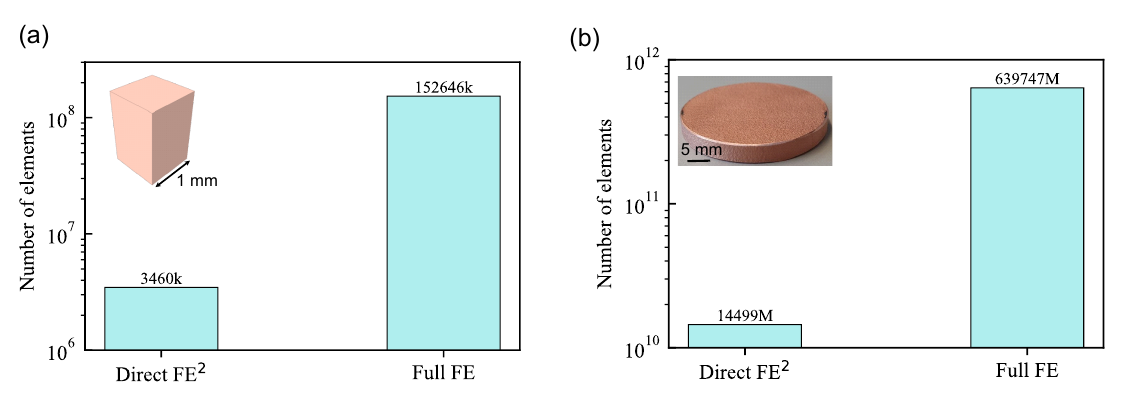}
    \caption{Comparison of the number of elements in the SPS geometry for the Direct FE$^2$ model and the full FE model for (a) the representative geometry and (b) an actual SPS-ed sample.}
    \label{fig:SPS_Element_Compare}
\end{figure}

To demonstrate the computational advantages of the Direct FE$^2$ method, a bar plot is presented in Fig. \ref{fig:SPS_Element_Compare}a. For an equisized geometry and mesh, the full FE model contains 44 times more elements than the Direct FE$^2$ model, based on a volumetric scaling factor of 3.533$^3$. As previously mentioned, the computational time required for the full FE model is at least 44 times greater than that for the Direct FE$^2$ model. 

Furthermore, in Fig. \ref{fig:SPS_Element_Compare}b, the extrapolated number of elements required for an actual SPS-ed sample is compared for the similar volumetric scaling factor. Simulating this directly with the full FE model is practically infeasible, but by applying an appropriate scaling factor, the number of elements—and consequently the computational cost—are significantly reduced in the Direct FE$^2$ model. It is important to note that the relationship between the DOF and computational time in FE solvers is often non-linear. For direct solvers, computational time is typically proportional to the cube of the number of DOF, which can lead to a substantial increase in processing time as the model size grows. Iterative solvers, while generally exhibiting a less severe scaling behavior, still tend to increase computational time super-linearly. 

\section{Conclusion}

We introduced an enhanced multiscale, multiphysics modeling framework for the spark plasma sintering (SPS) process using a novel Direct FE$^2$ method. This framework addresses critical limitations in existing models by integrating electro-thermal-mechanical effects at the macroscale with the granular behavior of powder particles at the microscale. 
The main findings of this work are:
\begin{itemize}
\item The proposed Direct FE$^2$ scheme effectively integrates power characteristics into electro-thermo-mechanical coupled problems with high accuracy, achieving a maximum error of less than 1\% in predicting displacement and temperature, closely matching the performance of the full FE solution.

\item Compared to the full FE method, the present Direct FE$^2$ method achieves acceleration ratios of up to 73 across various mesh sizes, significantly outperforming current computational techniques such as MPFEM and DEM in powder-level SPS simulations.

\item With its ability to accommodate diverse RVE configurations, the Direct FE$^2$ scheme demonstrates remarkable geometrical versatility and flexibility, provided all particles within the RVE are constrained. However, multi-particle boundary constraints may introduce irregularities due to unconstrained particles.

\item Densification rates are strongly influenced by both powder packing structures and current intensity, with FCC-type RVEs achieving higher overall densification despite slower rates due to their higher initial density. Additionally, lower currents, even under constant pressure, result in decreased densification rates, highlighting the crucial role of current intensity in the SPS process.

\end{itemize}

Future work should prioritize extending this method to unstructured macro meshes for better compatibility with complex geometries. Additionally, a promising direction for further research lies in developing generalized RVE formulations to account for non-spherical powder particles, advancing the realism and applicability of SPS simulations.
\section*{Acknowledgement}
The authors gratefully acknowledge the financial support of the Swiss National Science Foundation (SNSF) under Grant No. 200021\_200643. Special thanks are also extended to Marco Bernet for his valuable support in this work.

\appendix
\section{Johnson–Cook material model used for material characterization study}
\label{app:App_JC_model}
The yield stress of the matrix materials was assumed to follow the Johnson–Cook model \cite{johnson1985fracture} as
\begin{equation}
    \boldsymbol{\sigma_y} = (A + B \varepsilon^n) [1 + C ln(\varepsilon / \varepsilon_0)] \{ 1 - [(T-T_m) / (T_m - T_{room})]^m \}
\end{equation}
where $A$, $B$, $C$, $n$ and $m$ are material constants. $A$ is the initial yield stress, $B$ the hardening constant, $C$ the strain rate sensitivity index, $\varepsilon$ the strain rate ( the reference strain rate $\varepsilon_0$ = 1 $s^{-1}$), $n$ the hardening index, $m$ the temperature softening coefficient, $T_m$ the melting point and $T_{room}$ the room temperature.

\begin{table}
    \centering
     \caption{Material parameters of Cu for Johnson-Cook model \cite{dai2011numerical}}
    \begin{tabular}{l c c c c c c c} \hline 
         Material & $A$ (MPa)&  $B$ (MPa)&  $C$&  $n$&  $m$&  $T_{room}$ (K)& $T_m$ (K)\\ \hline 
         Cu & 90.0&  292&  0.025&  0.31&  1.09&  300& 1360\\ \hline
    \end{tabular}   
    \label{tab:Cu_Johnson_cook}
\end{table}

\section*{Data Availability}
The raw/processed data required to reproduce these findings can be shared upon reasonable request.




\bibliography{Bibliothek-dfe2}


\end{document}